\documentclass[aos,MSNbibl,seceqn,nameyear,rotating,dvips]{arximspdf}
\usepackage{algorithm}
\usepackage{multirow}
\usepackage{graphicx}

\doi{10.1214/14-AOS1223} %kopijuoti is PTS
\volume{42}
\issue{6}
\pubyear{2014}
\firstpage{2494}
\lastpage{2525}
\docsubty{FLA}

\makeatletter
\newcommand{\rrvert}{\vert}
\newcommand{\rrVert}{\Vert}
\newcommand{\llvert}{\vert}
\newcommand{\llVert}{\Vert}
\newtheorem{teo}{Theorem}[section]
\newtheorem{cor}{Corollary}
\newproclaim{mydef}{Definition}
\newproclaim{rem}{Remark}
\newcommand{\sgn}{\operatorname{sgn}}
\newcommand{\rnk}{\operatorname{rank}}
\newcommand{\argmin}{\arg\min}
\newcommand{\argmax}{\arg\max}
\newcommand{\mini}{\operatorname{minimize}\limits}
\makeatother

\begin{document}
\begin{frontmatter}

\title{Least quantile regression via modern optimization}
\runtitle{Least quantile regression}

\begin{aug}
\author[A]{\fnms{Dimitris}~\snm{Bertsimas}\ead[label=e1]{dbertsim@mit.edu}}
\and
\author[B]{\fnms{Rahul}~\snm{Mazumder}\corref{}\ead[label=e2]{rm3184@columbia.edu}}
\runauthor{D. Bertsimas and R. Mazumder}
\affiliation{Massachusetts Institute of Technology and Columbia University}
\address[A]{MIT Sloan School of Management\\
\quad and Operations Research Center\\
Massachusetts Institute of Technology\\
Cambridge, Massachusetts 02139\\
USA\\
\printead{e1}}
\address[B]{Department of Statistics\\
Columbia University\\
New York, New York 10027\\
USA\\
\printead{e2}}
\end{aug}

% HISTORY:
\received{\smonth{10} \syear{2013}}
\revised{\smonth{3} \syear{2014}}

% ABSTRACT
%
\begin{abstract}
We address the Least Quantile of Squares (LQS) (and in particular the
Least Median of Squares) regression problem using modern optimization
methods. We propose a Mixed Integer Optimization (MIO) formulation of
the LQS problem which allows us to find a provably global optimal
solution for the LQS problem.
Our MIO framework has the appealing characteristic that if we terminate
the algorithm early, we obtain a solution with a guarantee
on its sub-optimality.
We also propose continuous optimization methods based on first-order
subdifferential methods, sequential linear optimization and hybrid
combinations of them to obtain near optimal solutions to the LQS
problem. The MIO algorithm is found to benefit significantly from high
quality solutions delivered by our continuous optimization
based methods. We further show that the MIO approach leads to (a)
an optimal solution for \emph{any}
dataset, where the data-points $(y_{i}, \mathbf{x}_{i})$'s are not
necessarily in general position, (b) a simple proof of the
breakdown point of the LQS objective value that holds for any dataset
and (c) an extension to situations where there are
polyhedral constraints on the regression coefficient vector. We report
computational results with both synthetic and real-world datasets
showing that
the MIO algorithm with warm starts from the continuous optimization
methods solve small ($n=100$) and medium ($n=500$) size problems to
provable optimality in under two hours, and outperform all publicly
available methods for large-scale ($n={}$10,000) LQS problems.
\end{abstract}

% KEYWORDS
\end{frontmatter}

%s1 #&#
\section{Introduction}\label{secintro1}
Consider a linear model with response $\mathbf{y} \in\Re^{n}$, model
matrix $\mathbf{X}_{n \times p }$, regression coefficients
$\bolds\beta\in\Re^{p}$ and error ${\bolds\varepsilon} \in
\Re^{n}$:
\[
\mathbf{y} = \mathbf{X}\bolds\beta+ {\bolds\varepsilon}.
\]
We will assume that $\mathbf{X}$ contains a column of ones to account
for the intercept in the model.
Given data for the $i$th sample $(y_{i}, \mathbf{x}_{i})$, $i = 1,\ldots, n$ (where, $\mathbf{x}_{i} \in\Re^{p\times1}$) and
regression coefficients $\bolds\beta$, the
$i$th residual is given by the usual notation $r_{i} = y_{i} - \mathbf
{x}_{i}' \bolds\beta$ for $i = 1, \ldots, n$.
The traditional Least Squares (LS) estimator given by
%
%e1.1 #&#
\begin{equation}
\label{eqls} \hat{\bolds\beta}{}^{(\mathrm{LS})} \in\argmin _{\bolds\beta}
\sum_{i = 1}^{n} r_{i}^2
\end{equation}
is a popular and effective method for estimating the regression
coefficients when the error vector
${\bolds\varepsilon}$ has \emph{small} $\ell_2$-norm. However, in
the presence of outliers, the LS estimators do not work favorably---a
single outlier can have an arbitrarily large effect on the estimate.
The robustness of an estimator vis-a-vis outliers is often quantified
by the notion of
its finite sample breakdown point~[\citet{donoho1983notion,hampel1971general}].
% of an estimator.
% Donoho and Huber \cite{} introduced the finite sample notion of
%breakdown point of an estimator $\Theta(\M{y}, \M{X})$.
%Suppose $m$ among the $n$ sample points $(y_{i}, \M{x}_{i})$'s are
%changed by an arbitrary amount $(\Delta_{\M{y}}, \Delta_{\M{X}})$ to
%obtain
%the perturbed sample $(\M{y} + \Delta_{\M{y}}, \M{X} + \Delta_{
%Then, the breakdown point of the estimator $\Theta(y, \M{X})$ is
%given by the ratio $\widetilde{m}/n$, where $\widetilde{m}$
%denotes the minimum number of samples
%that need to be changed arbitrarily (with respect to $(\Delta_{\M{y}},
% $\Theta(\M{y} + \Delta_{\M{y}}, \M{X} + \Delta_{\M{X}}))$ becomes
%arbitrarily large.
The LS estimate~(\ref{eqls}) has a limiting (in the limit
$n\rightarrow\infty$ with $p$ fixed) breakdown point~[\citet
{hampel1971general}] of zero.

The Least Absolute Deviation (LAD) estimator given by
%
%e1.2 #&#
\begin{equation}
\label{eqlad} \hat{\bolds\beta}{}^{(\mathrm{LAD})} \in\argmin _{\bolds\beta}
\sum_{i = 1}^{n} \llvert r_{i}
\rrvert
\end{equation}
considers the
$\ell_{1}$-norm on the residuals, thereby implicitly assuming that the
error vector ${\bolds\varepsilon}$ has small
$\ell_{1}$-norm. The LAD estimator is not resistant to large
deviations in the covariates and, like the optimal LS solutions, has a
breakdown point of
zero (in the limit $n \rightarrow\infty$ with $p$ fixed).

M-estimators~[\citet{huber1973robust}] are obtained by minimizing
%$\ell_{1}$-norm on the residuals by a function
a loss function of the residuals of the form
$\sum_{i=1}^{n} \rho(r_{i})$, where $\rho(r)$ is a symmetric
function with a unique minimum at zero. Examples include the
Huber function and the Tukey function~[\citet{rousseeuw2005robust,huber2011robust}], among others. \mbox{M-}estimators often
simultaneously estimate the scale parameter along with the regression
coefficient.
M-estimators too are severely affected by the presence of outliers in
the covariate space.
A generalization of M-estimators are Generalized
\mbox{M-}estimators [\citet{rousseeuw2005robust,huber2011robust}],
which bound the influence of outliers in the covariate space by the
choice of a
weight function dampening the effect of outlying covariates. In some
cases, they have an improved
finite-sample breakdown point of $1/(p+1)$.

The repeated median estimator~[\citet{siegel1982robust}] with
breakdown point of approximately 50\%, was one of the earliest estimators
to achieve a very high breakdown point. The estimator however, is not
equivariant under linear transformations of the covariates.

\citet{rousseeuw1984least} introduced Least Median of Squares
(LMS)~[see also \citet{hampel1975beyond}] which
minimizes the median of the absolute residuals\footnote{Note that the
original definition of LMS~[\citet{rousseeuw1984least}] considers
the squared residuals instead of the absolute values. However, we will
work with the absolute values, since the problems are equivalent.}
%
%e1.3 #&#
\begin{equation}
\label{eqmedian1-lms} %\begin{myarray}{l c}
\hat{\bolds\beta}{}^{(\mathrm{LMS})} \in \argmin
_{\bolds\beta} \Bigl( \operatorname{median}\limits
_{i
= 1, \ldots, n } \llvert r_{i}
\rrvert \Bigr). %\end{myarray}
\end{equation}
The LMS problem is equivariant and has a limiting breakdown point of
50\%---making
it the first equivariant estimator to achieve the maximal possible
breakdown point in the limit $n\rightarrow\infty$ with $p$ fixed.

Instead of considering the median, one may consider more generally, the
$q$th order statistic, which leads to the
Least Quantile of Squares (LQS) estimator:
%
%e1.4 #&#
\begin{equation}
\label{eqmedian-q} \hat{\bolds\beta}{}^{(\mathrm{LQS})} \in \argmin _{\bolds\beta}
\llvert r_{(q)}\rrvert,
\end{equation}
where $r_{(q)}$ denotes the residual, corresponding to the $q$th
ordered absolute residual:
%
%e1.5 #&#
\begin{equation}
\label{order-elts-1} \llvert r_{(1)} \rrvert \leq\llvert r_{(2)}
\rrvert \leq\cdots\leq\llvert r_{(n)} \rrvert.
\end{equation}
\citet{rousseeuw1984least} showed that if the sample points
$(y_{i}, \mathbf{x}_{i})$, $i = 1, \ldots, n$ are in general position,
that is,
for any subset of ${\mathcal I} \subset\{1, \ldots, n \}$ with $\llvert
{\mathcal I} \rrvert  = p$, the $p\times p$ submatrix $X_{\mathcal I}$ has
rank $p$; an optimal LMS solution~(\ref{eqmedian1-lms}) exists and has
a finite sample breakdown point
of $(\lfloor n/2\rfloor- p + 2)/n$, where $\lfloor s\rfloor$ denotes
the largest integer smaller than or equal to $s$.
\citet{rousseeuw1984least} showed that the finite sample
breakdown point of the estimator~(\ref{eqmedian1-lms})
can be further improved to achieve the maximum possible finite sample
breakdown point
if one considers the estimator~(\ref{eqmedian-q}) with $q = \lfloor
n/2 \rfloor+ \lfloor(p +1)/2 \rfloor$.
The LMS estimator has low efficiency~[\citet{rousseeuw1984least}]. This can, however, be improved by using certain
post-processing methods on the LMS estimator---the one step M-estimator
of~\citet{bickel1975one} or
a reweighted least-squares estimator, where points with large values of
LMS residuals are given small weight are popular methods that are used
in this vein.

\subsection*{Related work}
It is a well recognized fact that the LMS problem is computationally
demanding due to the combinatorial nature of the problem.
\citet{bernholt-hard} showed that computing an optimal LMS
solution is NP-hard.

Many algorithms based on different approaches have been proposed for
the LMS problem over the past thirty years.
State of the art algorithms, however, fail to obtain a global minimum of
the LMS problem for problem sizes larger than $n = 50, p = 5$. This
severely limits the use of LMS for important real world multivariate
applications,
where $n$ can easily range in the order of a few thousands.
It goes without saying that a poor local minimum for the LMS problem
may be misleading from a statistical inference point of view [see also
\citet{Stromberg1993} and references therein for related
discussions on this matter].
The various algorithms presented in the literature for the LMS can be
placed into two very broad categories.
One approach computes an optimal solution to the LMS problem using
geometric characterizations of the fit---they typically rely on
complete enumeration and have complexity $O(n^p)$.
The other approach gives up on obtaining an
optimal solution and resorts to heuristics and/or randomized algorithms
to obtain approximate solutions to the LMS problem. These methods,
to the best of our knowledge, do not provide certificates about the
quality of the solution obtained. We describe below a brief overview of
existing algorithms for LMS.

Among the various algorithms proposed in the literature for the LMS
problem, the most popular seems to be
{\small\textsc{PROGRESS}} (Program for Robust Regression)~[\citet
{rousseeuw2005robust,Rousseeuw97recentdevelopments}]. The \mbox{algorithm}
does a complete enumeration of all $p$-subsets of the $n$ sample
points, computes the hyperplane passing through them and finds the
configuration leading to the smallest value of the objective. The
algorithm has a run-time complexity of $O(n^p)$ and assumes that
the data points are in general position.
For computational scalability, heuristics that randomly sample subsets
are often used.
See also~\citet{Barreto2006SCL16479631648203} for a recent work
on algorithms for the bivariate regression problem.

%Approximation algorithms and Randomized Algorithms have also been
%proposed for the problem (see for example the works of ~\cite{}).
\citet{steele1986algorithms} proposed exact algorithms for LMS
for $p = 2$ with complexity $O(n^3)$ and some probabilistic speed-up methods
with complexity $O((n\log(n))^2)$.

\citet{Stromberg1993} proposed an exact algorithm for LMS with
run-time $O(n^{(p+2)}\log(n))$ using some
insightful geometric properties of the LMS fit. This method does a
brute force search
among ${n \choose p+1}$ different regression coefficient values and
scales up to problem sizes $n=50$ and $p=5$.

\citet{agullo-1997} proposed a finite branch and bound technique
with run-time complexity $O(n^{p+2})$ to obtain an optimal solution to
the LMS problem motivated
by the work of~\citet{Stromberg1993}. The algorithm showed
superior performance compared to methods preceding it
and can scale up to problem sizes $n \approx70, p \approx4$.

\citet{erickson2006least} give an exact algorithm with run-time
$O(n^p\log(n))$ for LMS
and also show that computing an optimal LMS solution requires $O(n^p)$ time.
For the two-dimensional case $p = 2$, \citet{guided-topo-jasa-87}
proposed an exact algorithm for LMS with complexity
$O(n^2)$ using the topological sweep-line technique.

\citet{Giloni2002LTS22581142258762} propose integer optimization
formulations for the LMS problem, however,
no computational experiments are reported---the practical performance
of the proposed method thus remains unclear.

\citet{Mount20072461} present an algorithm based on branch and
bound for $p=2$ for computing approximate solutions to the LMS problem.
\citet{mount-quantile} present a quantile approximation algorithm
with approximation factor $\varepsilon$ with complexity
$O(n\log(n) + (1/\varepsilon)^{O(p)})$.
\citet{chakraborty2008optimization} present probabilistic search
algorithms for a class of problems in robust statistics.
\citet{Nunkesser20103242} describe computational procedures based
on heuristic search strategies using evolutionary algorithms for some
robust statistical estimation problems including LMS.
\citet{Hawkins199381} proposes a probabilistic algorithm for LMS
known as the ``Feasible Set Algorithm'' capable of
solving problems up to sizes $n = 100, p = 3$.

\citet{Bernholt05computingthe} describes a randomized algorithm
for computing the LMS running in
$O(n^p)$ time and $O(n)$ space, for fixed $p$.
\citet{Olson97anapproximation} describes an approximation
algorithm to compute an optimal LMS solution within an approximation
factor of two using randomized
sampling methods---the method has (expected) run-time complexity of
$O(n^{p-1}\log(n))$.

\subsection*{Related approaches in robust regression}
Other estimation procedures that achieve a high breakdown point and
good statistical efficiency
include the least trimmed squares estimator~[\citet
{rousseeuw1984least,rousseeuw2005robust}], which minimizes the sum of
squares of the $q$ smallest squared residuals. Another popular approach
is based on S-estimators~[\citet
{rousseeuw1984least,rousseeuw2005robust}], which are
a type of M-estimators of scale on the residuals.
These estimation procedures like the LMS estimator are
NP-hard~[\citet{bernholt-hard}].

We refer the interested reader to~\citet{hubert2008high} for a
nice review of various robust statistical methods and
their applications~[\citet
{meer1991robust,comp-vision-siam-review-99,rousseeuw2006robustness}].

\subsection*{What this paper is about}
In this paper, we propose a computationally tractable framework to
compute a globally optimal solution to the LQS problem~(\ref
{eqmedian-q}), and in particular the LMS problem
via modern optimization methods: first-order methods from continuous
optimization and mixed integer optimization (MIO), see~\citet
{bertsimas2005optimization}.
Our view of computational tractability is not polynomial time solution
times as these do not exist for the LQS problem unless P${}={}$NP. Rather it
is the ability of a method to solve problems of practical interest in
times that are appropriate for the application addressed.
An important advantage of our framework is that it easily adapts to
obtain solutions to more general variants of~(\ref{eqmedian-q}) under
polyhedral constraints, that is,
%
%e1.6 #&#
\begin{equation}
\label{eqmedian-q-abs-gen} \mini_{\bolds\beta}\qquad \llvert r_{(q)} \rrvert,\qquad
\mbox{subject to}\qquad \mathbf{A}\bolds\beta\leq\mathbf{b}, %\begin{array}{l l c}
\end{equation}
where $\mathbf{A}_{m \times p}, \mathbf{b}_{m \times1}$ are given
parameters in the problem representing
side constraints on the variable $\bolds\beta$ and ``$\leq$''
denotes component wise inequality.
This is useful if one would like to incorporate some form of
regularization on the $\bolds\beta$ coefficients, for example,
$\ell_{1}$ regularization~[\citet{Ti96}] or generalizations thereof.
% or a generalized\footnote{A generalized $\ell_{1}$ regularization on
%the regression coefficients is given by
%a constraint set of the form $\left\| D\B\beta\left\| _{1} \leq\lambda$ for some
%given matrix $D_{m \times p}$ and a shrinkage/sparsity level $
%$\ell_{1}$ regularization on $\B\beta$~\citep{ryan-genlasso}.

\subsection*{Contributions}
Our contributions in this paper may be summarized as follows:
\begin{longlist}[(3)]
\item[(1)] We use MIO to find a provably optimal solution to the LQS
problem. Our
framework has the appealing characteristic that if we terminate the
algorithm early, we obtain a solution with a guarantee
on its suboptimality. We further show that the MIO approach leads to an
optimal solution for \emph{any}
dataset where the data-points $(y_{i}, \mathbf{x}_{i})$'s are not
necessarily in general position.
Our framework enables us to provide a simple proof of the breakdown
point of the LQS objective value, generalizing the existing results for
the problem.
Furthermore, our approach is readily generalizable to
problems of the type~(\ref{eqmedian-q-abs-gen}).

\item[(2)] We introduce a variety of solution methods based on modern
continuous optimization---first-order
subdifferential based minimization, sequential linear optimization and
a hybrid version of these two methods that provide near optimal
solutions for
the LQS problem. The MIO algorithm is found to significantly benefit
from solutions obtained by the continuous optimization methods.

\item[(3)] We report computational results with both synthetic and
real-world datasets that show that
the MIO algorithm with warm starts from the continuous optimization
methods solve small ($n=100$) and medium ($n=500$) size LQS problems to
provable optimality in under two hours, and outperform all publicly
available methods for large-scale ($n={}$10,000) LQS problems, but
without showing provable optimality in under two hours of computation time.
\end{longlist}

\subsection*{Structure of the paper}
The paper is organized as follows. Section~\ref{secmio-method1}
describes MIO approaches for the LQS problem.
Section~\ref{secconts-opt-methods1} describes continuous optimization
based methods for obtaining
local minimizers of the LQS problem.
Section~\ref{secprops-lqs} describes properties of an optimal LQS
solution. Section~\ref{seccomps-1} describes computational
results and experiments. The last section contains our key conclusions.

%s2 #&#
\section{Mixed integer optimization formulation}\label{secmio-method1}
In this section, we present an exact MIO formulation for the LQS problem.
For the sake of completeness, we will first introduce the definition of
a linear MIO problem.
The generic MIO framework concerns the following optimization problem:
%
%e2.1 #&#
\begin{eqnarray}\label{mio-gen-setup}
\mini &\qquad& \mathbf{c}'\bolds{\alpha} + \mathbf{d}' \bolds{\theta},\nonumber
\\
&&A\bolds{\alpha} + B\bolds\theta\bolds\geq \mathbf{b},
\nonumber\\[-8pt]\\[-8pt]\nonumber
&&\bolds\alpha\in\Re^n_{+},
\\
&&\bolds\theta\in\{0, 1\}^m,\nonumber
\end{eqnarray}
where $\mathbf{c}\in\Re^{n}, \mathbf{d}\in\Re^{m}, A \in\Re^{k
\times n}, B \in\Re^{k \times m}, \mathbf{b} \in\Re^{k}$ are the
given parameters of the problem;
$\Re^{n}_{+}$ denotes the nonnegative $n$-dimensional orthant, the
symbol $\bolds{\geq}$ denotes element-wise inequalities and
we optimize over both continuous ($\bolds{\alpha}$) and discrete
($\bolds\theta$) variables. For background on MIO,
see~\citet{bertsimas2005optimization}.

Consider a list of $n$ numbers $\llvert r_{1}\rrvert, \ldots, \llvert r_{n}\rrvert $, with the
ordering described in~(\ref{order-elts-1}).
To model the sorted $q$th residual, that is, $\llvert r_{(q)}\rrvert $, we need to
express the fact that $r_{i} \leq\llvert r_{(q)}\rrvert $
for $q$ many residuals $\llvert r_{i}\rrvert $'s from $\llvert r_{1}\rrvert, \ldots, \llvert r_{n}\rrvert $. To
do so, we introduce the binary variables $z_{i}$, $i = 1, \ldots, n$
with the interpretation
%
%e2.2 #&#
\begin{equation}
\label{zi-def-1} z_{i} = \cases{ 1, &\quad if $\llvert r_{i}
\rrvert \leq\llvert r_{(q)}\rrvert $,
\cr
0, &\quad otherwise.}
\end{equation}
We further introduce auxiliary continuous variables $\mu_{i},
\bar{\mu}_{i} \geq0$, such that
%
%e2.3 #&#
\begin{equation}
\label{order-resid-1} \llvert r_{i}\rrvert - \mu_{i} \leq\llvert
r_{(q)} \rrvert \leq\llvert r_{i}\rrvert + \bar{\mu
}_{i},\qquad i = 1, \ldots, n,
\end{equation}
with the conditions
%
%e2.4 #&#
\begin{eqnarray}
\label{eqnmus-1}
\mbox{if } \llvert r_{i}\rrvert &\geq& \llvert
r_{(q)}\rrvert,\qquad\mbox{then }\bar{\mu}_{i}=0,
\mu_{i} \geq0\quad\mbox{and}
\nonumber
\\[-8pt]
\\[-8pt]
\nonumber
\mbox{if } \llvert r_{i}\rrvert &\leq& \llvert
r_{(q)}\rrvert,\qquad\mbox{then } \mu_{i}=0, \bar{\mu}_{i} \geq0.
\end{eqnarray}

We thus propose the following MIO formulation:
%
%e2.5 #&#
\begin{eqnarray}\label{lqs-obs-1}
&& \mini\qquad \gamma,\nonumber
\\
&&\qquad\mbox{subject to}\qquad \llvert r_{i}\rrvert + \bar{
\mu}_{i} \geq\gamma,\qquad i = 1,\ldots, n,\nonumber
\\
&&\phantom{\qquad\mbox{subject to}\qquad}\gamma\geq\llvert r_{i}\rrvert - \mu_{i},\qquad i = 1,\ldots, n,\nonumber
\\
&&\phantom{\qquad\mbox{subject to}\qquad}M_{u}z_{i} \geq\bar{\mu}_{i},\qquad i = 1,\ldots, n,\nonumber
\\
&&\phantom{\qquad\mbox{subject to}\qquad}M_{\ell} ( 1- z_{i}) \geq\mu_{i},\qquad i = 1,\ldots, n,
\\
&&\phantom{\qquad\mbox{subject to}\qquad}\sum_{i=1}^{n} z_{i} = q,\nonumber
\\
&&\phantom{\qquad\mbox{subject to}\qquad}\mu_{i} \geq0,\qquad i = 1, \ldots, n,\nonumber
\\
&&\phantom{\qquad\mbox{subject to}\qquad}\bar{\mu}_{i} \geq0,\qquad i = 1, \ldots, n,\nonumber
\\
&&\phantom{\qquad\mbox{subject to}\qquad}z_{i} \in\{0, 1\},\qquad i = 1, \ldots, n,\nonumber
\end{eqnarray}
where, $\gamma, z_{i},\mu_{i}, \bar{\mu}_{i}$, $i = 1, \ldots,n$ are the optimization variables,
$M_{u}, M_{\ell}$ are the so-called \emph{Big-M} constants.
Let us denote the optimal solution of problem~(\ref{lqs-obs-1}), which
depends on $M_{\ell}, M_{u}$,
by $\gamma^*$.
Suppose we consider $M_{u}, M_{\ell} \geq\max_{i} \llvert r_{(i)}\rrvert $---it
follows from formulation~(\ref{lqs-obs-1}) that
$q$ of the $\mu_{i}$'s are zero. Thus, $\gamma^*$ has to be larger
than at least $q$ of the $\llvert r_{i}\rrvert $ values.
By arguments similar to the above, we see that,
since $(n - q)$ of the $z_{i}$'s are zero, at least $(n - q)$ many
$\bar{\mu}_{i}$'s are zero. Thus,
$\gamma^{*}$ is less than or equal to at least $(n - q)$ many of the
$\llvert r_{i}\rrvert, i=1, \ldots, n$ values. This shows that
$\gamma^{*}$ is indeed equal to $\llvert r_{(q)}\rrvert $, for $M_{u}, M_{\ell}$
sufficiently large.

We found in our experiments that, in formulation~(\ref{lqs-obs-1}), if
$z_{i}=1$, then $\bar{\mu}_{i}=M_{u}$ and if $z_{i}= 0$ then
$\mu_{i}=M_{\ell}$.
Though this does not interfere with the definition of~$\llvert r_{(q)}\rrvert $, it
creates a difference in the strength of the MIO formulation.
%The constraint~\eqref{sum-bound1} is indeed natural given the
%representation~\eqref{order-resid-1} and observing that the
We describe below how to circumvent this shortcoming.

From~(\ref{eqnmus-1}), it is clear that $\bar{\mu}_{i}\mu_{i}
=0$, $\forall i = 1, \ldots, n$.
%i.e., both the
%variables $\mu_{i}$ and $\bar{\mu}_{i}$ cannot be nonzero at the
%same time.
The constraint $\bar{\mu}_{i}\mu_{i} =0$ can be modeled via
integer optimization using Specially Ordered Sets of type~1 [\citet{bertsimas2005optimization}], that is, SOS-1 constraints
as follows:
%
%e2.6 #&#
\begin{equation}
\label{sos-def-1} \mu_{i}\bar{\mu}_{i}=0\quad\iff\quad (\mu_{i}, \bar{\mu}_{i} ) \dvtx  \mbox{SOS-1},
\end{equation}
for every $i = 1, \ldots, n$.
In addition, observe that, for $M_{\ell}$ sufficiently large and every
$i\in\{1, \ldots, n\}$
the constraint $M_{\ell} ( 1- z_{i}) \geq\mu_{i} \geq0$ can be
modeled\footnote{To see why this is true, observe that
$(\mu_{i}, z_{i}) \dvtx  \mbox{SOS-1}$ is equivalent to $\mu_{i}z_{i}
=0$. Now, since \mbox{$z_{i} \in\{ 0, 1\}$}, we have the following possibilities:
$z_{i}= 0$, in which case $\mu_{i}$ is free; if $z_{i} = 1$, then $\mu
_{i} = 0$. }
by a SOS-1 constraint---$(\mu_{i}, z_{i}) \dvtx  \mbox{SOS-1}$.
% It also follows from~\eqref{lq-obs-1} that $z_{i} \mu_{i} = 0, i =
%1, \ldots, n$---which can be modeled by SOS-1
%constraints i.e.$( \mu_{i}, z_{i} ) \dvtx  \mbox{SOS-1}$ for $i = 1,
In light of this discussion, we see that
%
%e2.7 #&#
\begin{equation}
\label{eqn-sos-split1} \llvert r_{i}\rrvert - \llvert r_{(q)}\rrvert
= {\mu}_{i} - \bar{\mu}_{i},\qquad ( \mu _{i},
\bar{\mu}_{i} ) \dvtx  \mbox{SOS-1}.
\end{equation}
We next show that $\llvert  r_{(q)}\rrvert  \geq\bar{\mu}_{i}$ and $\mu_{i}
\leq\llvert r_{i}\rrvert $ for all $i = 1, \ldots, p$.
When $\llvert  r_{i}\rrvert  \leq\llvert  r_{(q)}\rrvert $, it follows from the above
representation that
\[
{\mu}_{i} =0 \quad\mbox{and}\quad\bar{\mu}_{i} =
\llvert r_{(q)}\rrvert - \llvert r_{i}\rrvert \leq\llvert
r_{(q)}\rrvert.
\]
When $\llvert  r_{i}\rrvert  > \llvert  r_{(q)}\rrvert $, it follows that $\bar{\mu}_{i} =0$.
Thus, it follows that $0 \leq\bar{\mu}_{i} \leq\llvert  r_{(q)}\rrvert $ for
all $i = 1, \ldots, n$.
It also follows by a similar argument that
$0 \leq\mu_{i} \leq\llvert  r_{i}\rrvert $ for all $i$.

%In addition, observe that, for $M_{\ell}$ sufficiently large and every
%$i\in\{1, \ldots, n\}$
%the constraint $M_{\ell} ( 1- z_{i}) \geq\mu_{i} \geq0$, can be
%equivalently expressed
%by a SOS-1 constraint---$(\mu_{i}, z_{i}) \dvtx  \mbox{SOS-1}$. To see why
%this is true observe that
%$(\mu_{i}, z_{i}) \dvtx  \mbox{SOS-1}$ is equivalent to $\mu_{i}z_{i} =0$.
%Now, since $z_{i} \in\{ 0, 1\}$, we have the following possibilities\dvtx
%$z_{i}= 0$, in which case $\mu_{i}$ is free; if $z_{i} = 1$, then $

%In light of the above discussion, we present below the following MIO
%formulation for the LQS estimator
Thus, by using $\mbox{SOS-1}$ type of constraints, we can avoid the
use of \emph{Big-M}'s appearing in formulation~(\ref{lqs-obs-1}), as follows:
%
%e2.8 #&#
\begin{eqnarray}\label{lqs-obs-1-mod}
&&\mini\qquad \gamma,\nonumber
\\
&&\qquad\mbox{subject to}\qquad \llvert r_{i}\rrvert - \gamma=\mu_{i} - \bar{\mu }_{i},\qquad i = 1,\ldots, n,\nonumber
\\
&&\phantom{\qquad\mbox{subject to}\qquad}\sum_{i=1}^{n} z_{i} = q,\nonumber
\\
&&\phantom{\qquad\mbox{subject to}\qquad}\gamma\geq\bar{\mu}_{i},\qquad i = 1,\ldots, n,\nonumber
\\
&&\phantom{\qquad\mbox{subject to}\qquad}\bar{\mu}_{i} \geq0, \qquad i = 1,\ldots, n,
\\
&&\phantom{\qquad\mbox{subject to}\qquad}\mu_{i}\geq0, \qquad i = 1, \ldots, n,\nonumber
\\
&&\phantom{\qquad\mbox{subject to}\qquad}(\bar{\mu}_{i}, \mu_{i}) \dvtx  \mbox{SOS-1}, \qquad i = 1, \ldots, n,\nonumber
\\
&&\phantom{\qquad\mbox{subject to}\qquad}( z_{i}, {\mu}_{i} ) \dvtx \mbox{SOS-1},\qquad i = 1,\ldots, n,\nonumber
\\
&&\phantom{\qquad\mbox{subject to}\qquad}z_{i} \in\{0, 1\},\qquad i = 1, \ldots, n.\nonumber
\end{eqnarray}
Note, however, that the constraints
%
%e2.9 #&#
\begin{equation}
\label{const-1} \llvert r_{i}\rrvert - \gamma= \mu_{i} -
\bar{\mu}_{i},\qquad i = 1, \ldots, n
\end{equation}
are not convex in $r_{1}, \ldots, r_{n}$.
We thus introduce the following variables ${r}_{i}^{+}, {r}_{i}^{-}$, $i
= 1, \ldots, n$ such that
%
%e2.10 #&#
%e2.11 #&#
\begin{eqnarray}\label{resids-split-1}
r_{i}^{+} + r_{i}^{-}= \llvert r_{i}\rrvert,\qquad
y_{i} - \mathbf{x}_{i}'\bolds\beta= r_{i}^{+} -r_{i}^{-},
\nonumber\\[-8pt]\\[-8pt]
\eqntext{r^+_{i} \geq0, r^-_{i}\geq0, r^+_{i}r^-_{i}=0, i =1, \ldots, n.}
\end{eqnarray}
The constraint $r^+_{i}r^-_{i}=0$ can be modeled via SOS-1 constraints
\[
\bigl( r^+_{i}, r^-_{i} \bigr) \dvtx  \mbox{SOS-1}\qquad
\mbox{for every } i = 1, \ldots, n.
\]

\begin{figure}%[!h]

\includegraphics{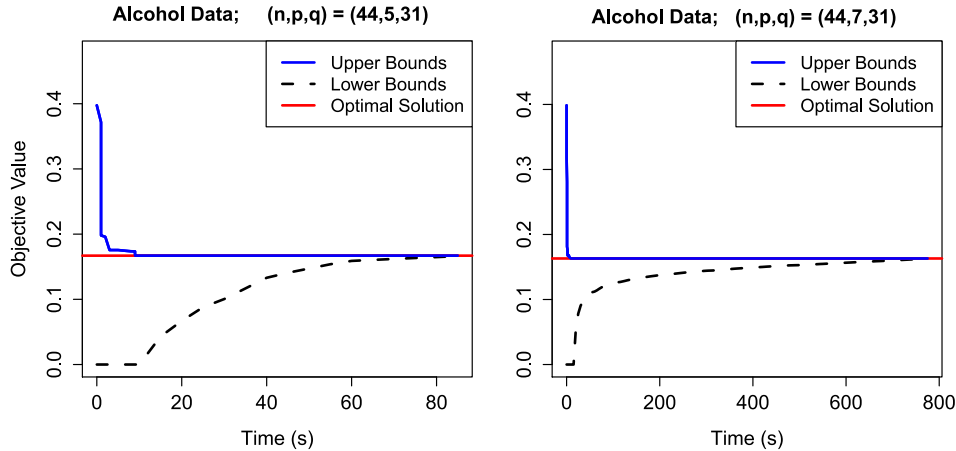}

\caption{Figure showing the typical evolution of the MIO
formulation~(\protect\ref{lqs-reg-form-1}) for the ``Alcohol''
dataset with $n = 44, q = 31$ with $p=5$ (left panel) and
$p = 7$ (right panel). Global solutions for both the problems are found
quite quickly in both examples, but it takes longer to certify global
optimality via the lower bounds.
As expected, the time taken for the MIO to certify convergence to the
global optimum increases with increasing $p$.}\label{fig-alco-data1}
\end{figure}

This leads to the following MIO for the LQS problem that we use in this paper:
%
%e2.12 #&#
\begin{eqnarray}\label{lqs-reg-form-1}
&&\mini\qquad \gamma,\nonumber
\\
&&\qquad\mbox{subject to}\qquad r_{i}^{+} + r_{i}^{-}- \gamma= \bar{\mu }_{i} - \mu_{i}, \qquad i = 1,\ldots, n,\nonumber
\\
&&\phantom{\qquad\mbox{subject to}\qquad} r_{i}^{+} - r_{i}^{-} =y_{i} - \mathbf{x}_{i}'\bolds\beta, \qquad i =1,\ldots, n,\nonumber
\\
&&\phantom{\qquad\mbox{subject to}\qquad} \sum_{i=1}^{n} z_{i} = q,
\\
&&\phantom{\qquad\mbox{subject to}\qquad}\gamma\geq\mu_{i} \geq0, \qquad i = 1,\ldots, n,\nonumber
\\
&&\phantom{\qquad\mbox{subject to}\qquad}\mu_{i} \geq0, \qquad i = 1,\ldots, n,\nonumber
\\
&&\phantom{\qquad\mbox{subject to}\qquad} \bar{\mu}_{i}\geq0, \qquad i = 1, \ldots, n,\nonumber
\\
&&\phantom{\qquad\mbox{subject to}\qquad} r_{i}^{+} \geq0, r_{i}^{-} \geq0,\qquad i = 1,\ldots, n,\nonumber
\\
&&\phantom{\qquad\mbox{subject to}\qquad} (\bar{\mu}_{i}, \mu_{i}) \dvtx  \mbox{SOS-1},\qquad i =1, \ldots, n,\nonumber
\\
&&\phantom{\qquad\mbox{subject to}\qquad} \bigl( r^+_{i}, r^-_{i} \bigr) \dvtx  \mbox{SOS-1},\qquad i= 1,\ldots, n,\nonumber
\\
&&\phantom{\qquad\mbox{subject to}\qquad} ( z_{i}, {\mu}_{i} ) \dvtx \mbox{SOS-1},\qquad i = 1, \ldots,n,\nonumber
\\
&&\phantom{\qquad\mbox{subject to}\qquad} z_{i} \in\{0, 1\}, \qquad i = 1, \ldots, n.\nonumber
\end{eqnarray}
To motivate the reader, we show in Figure~\ref{fig-alco-data1} an
example that illustrates that the MIO formulation~(\ref
{lqs-reg-form-1}) leads to a provably optimal solution
for the LQS problem. We give more details in Section~\ref{seccomps-1}.
%
%f1 #&#
%
%% left, bottom, right, top

%s3 #&#
\section{Continuous optimization based methods} \label{secconts-opt-methods1}
We describe two main approaches based on continuous optimization for
the LQS problem. Section~\ref{secseq-LP1} presents a method based on
sequential linear optimization and Section~\ref{secsubgrad1} describes a
first-order subdifferential based method for the LQS problem.
Section~\ref{sechybrid-1} describes hybrid combinations of the
aforementioned approaches,
which we have found, empirically, to provide high quality solutions.
Section~\ref{secinit-strategies} describes initialization strategies
for the algorithms.

%s3.1 #&#
\subsection{Sequential linear optimization}\label{secseq-LP1}
We describe a sequential linear optimization approach to obtain a local
minimum of problem~(\ref{eqmedian-q}).
We first describe the algorithm, present its convergence analysis and
describe its iteration complexity.

\subsubsection*{Main description of the algorithm}\label{secset-up-1}
%The proposal relies on the following characterization of the $q$th
%ordered absolute residual:
We decompose the $q$th ordered absolute residual as follows:
%
%e3.1 #&#
\begin{equation}
\label{eqmedian2} \llvert r_{(q)}\rrvert = \bigl\llvert y_{(q)} -
\mathbf{x}'_{(q)}\bolds\beta\bigr\rrvert = \underbrace{\sum
_{i= q }^{n} \bigl\llvert y_{(i)} -
\mathbf{x}'_{(i)}\bolds\beta \bigr\rrvert }_{H_{q}(\bolds\beta)} -
\underbrace{\sum_{i=q+1}^{n} \bigl\llvert
y_{(i)} - \mathbf{x}'_{(i)}\bolds\beta\bigr\rrvert
}_{H_{q+1}(\bolds\beta)},
\end{equation}
where, we use the notation $H_{m}(\bolds\beta) = \sum_{i=
m}^{n} \llvert y_{(i)} - \mathbf{x}'_{(i)}\bolds\beta\rrvert $ to denote the
sum of the largest $m$ ordered
residuals $\llvert r_{(i)}\rrvert:=\llvert y_{(i)} - \mathbf{x}'_{(i)}\bolds\beta\rrvert$,
$i = 1, \ldots, n$ in absolute value.
%We will show that the functions $H_{q}(\B\beta) and $H_{q+1}(\B\beta)$
%are both convex in $\B\beta$.
%We will show that the function $H_{m}(\B\beta)$ convex in $\B\beta$
%for any $m \in\{ 1, \ldots, n \}$. Towards this end
The function $H_{m}(\bolds\beta)$
can be written as
%
%e3.2 #&#
\begin{eqnarray}\label{top-r-convex-1}
&& H_{m}(\bolds\beta):= \max_{\mathbf{w}}\qquad
\sum_{i=1}^{n} w_{i} \bigl\llvert
y_{i} - \mathbf{x}'_{i}\bolds\beta\bigr\rrvert\nonumber
\\
&&\qquad \mbox{subject to}\qquad \sum_{i=1}^{n} w_{i} = n-m +1,
\\
&&\phantom{\qquad \mbox{subject to}\qquad} 0 \leq w_{i} \leq1,\qquad i = 1, \ldots, n.\nonumber
\end{eqnarray}
Let us denote the feasible set in problem~(\ref{top-r-convex-1}) by
\[
{\mathcal W}_{m}:= \Biggl\{ \mathbf{w} \dvtx  \sum
_{i=1}^{n} w_{i} = n-m +1,
w_{i} \in[0, 1], i = 1, \ldots, n \Biggr\}.
\]
Observe that for every $\mathbf{w}\in{\mathcal W}_{m}$ the function
$\sum_{i=1}^{n} w_{i} \llvert y_{i} - \mathbf{x}'_{i}\bolds\beta\rrvert $ is
convex in $\bolds\beta$. Furthermore, since
$H_{m}(\bolds\beta)$ is the point-wise supremum
with respect to $\mathbf{w}$ over~${\mathcal W}_{m}$, the function
$H_{m}(\bolds\beta)$ is convex in $\bolds\beta$
[see~\citet{BV2004}].
Equation~(\ref{eqmedian2}) thus shows that $\llvert r_{(q)}\rrvert $ can be written as the
difference of two convex functions, namely,
$H_{q}(\bolds\beta)$ and $H_{q+1}(\bolds\beta)$.
By taking the dual of problem~(\ref{top-r-convex-1}) and invoking
strong duality, we have
%
%e3.3 #&#
\begin{eqnarray}\label{lp-duality-1}
&& H_{m}(\bolds\beta) = \min_{\theta,\bolds\nu}\qquad \theta (n-m+1) + \sum_{i=1}^{n}\nu_{i}\nonumber
\\
&&\qquad\mbox{subject to}\qquad \theta+ \nu_{i} \geq\bigl\llvert y_{i} - \mathbf {x}'_{i}\bolds\beta\bigr\rrvert,\qquad i = 1, \ldots,n,
\\
&&\phantom{\qquad\mbox{subject to}\qquad}\nu_{i} \geq0,\qquad i = 1, \ldots, n.\nonumber
\end{eqnarray}
Representation~(\ref{top-r-convex-1}) also provides
a characterization of the set of subgradients of~$H_{m}(\bolds\beta)$:
%
%e3.4 #&#
\begin{eqnarray}
\label{subdiff-top-r-conv} && {\large{\bolds\partial} H_{m}(\bolds\beta)}
\nonumber
\\[-8pt]
\\[-8pt]
\nonumber
&&\qquad =
\operatorname{conv} \Biggl\{ \sum_{i=1}^{n}-
w^*_{i} \sgn\bigl(y_{i} - \mathbf{x}'_{i}
\bolds\beta\bigr)\mathbf{x}_{i} \dvtx  \mathbf{w}^* \in\argmax_{\mathbf{w} \in{\mathcal W}_{m} }
{\mathcal L}(\bolds\beta, \mathbf{w}) \Biggr\},
\end{eqnarray}
where ${\mathcal L}(\bolds\beta, \mathbf{w}) = \sum_{i=1}^{n}
w_{i} \llvert y_{i} - \mathbf{x}'_{i}\bolds\beta\rrvert $ and ``conv($S$)''
denotes the convex hull of set~$S$.
An element of the set of subgradients (\ref{subdiff-top-r-conv}) will
be denoted by $\partial H_{m} (\bolds\beta)$.

%Since $H_{q}(\B\beta)$ and $H_{q+1}(\B\beta)$ are both convex in $\B
% Representation~\eqref{eqmedian2} shows that the
% LQS can be written as the difference of two convex functions.
%Before we go into the details of the algorithm, we describe the main
%idea involved.
%It turns out that $H_{q}(\B\beta)$ can be written as linear
%optimization problem.
%
% Having expressed $\left|r_{q}\right|$ as the difference of two convex functions,
%we next describe the sequential linear optimization (LO) algorithm.
%the main idea of the sequential LO algorithm.
% We use a sequential linear optimization method to minimize Problem~

Recall that~(\ref{eqmedian2}) expresses the $q$th ordered absolute
residual as a difference of two convex functions. Now,
having expressed $H_{q}(\bolds\beta)$ as the value of a Linear Optimization (LO)
problem~(\ref{lp-duality-1}) (with $m=q$) we linearize the function
$H_{q+1}(\bolds\beta)$.
If $\bolds\beta_{k}$ denotes the value of the estimate at
iteration $k$,
we linearize $H_{q+1}(\bolds\beta)$ at $\bolds\beta_{k}$
as follows:
%
%e3.5 #&#
\begin{equation}
\label{eqapprox-1aa} H_{q+1}(\bolds\beta) \approx H_{q+1}(\bolds
\beta_{k} ) + \bigl\langle\partial H_{q+1}(\bolds
\beta_{k} ), \bolds\beta- \bolds\beta_{k} \bigr\rangle,
\end{equation}
where $\partial H_{q+1}(\bolds\beta_{k} )$ is a subgradient of
$H_{q+1}(\bolds\beta_{k} )$ as defined in~(\ref{subdiff-top-r-conv}), with \mbox{$m = (q+1)$}.

Combining (\ref{lp-duality-1}) and (\ref{eqapprox-1aa}), we obtain
that, the minimum of problem~(\ref{eqmedian2}) with respect to
$\bolds\beta$ can be approximated by
%Problem~\eqref{eqapprox-1-update} is equivalent to
solving the following LO problem:
%
%e3.6 #&#
\begin{eqnarray}\label{lqs-lp-almost-20}
&& \min_{\bolds{\nu}, \theta, \bolds\beta}\qquad \theta( n - q +1 ) + \sum
_{i=1}^{n} \nu_{i} - \bigl\langle\partial H_{q+1}(\bolds \beta_{k}), \bolds\beta\bigr\rangle\nonumber
\\
&&\qquad\mbox{subject to}\qquad \theta+ \nu_{i} \geq\bigl\llvert
y_{i} - \mathbf {x}'_{i}\bolds\beta\bigr\rrvert, \qquad i = 1, \ldots,n,
\\
&&\phantom{\qquad\mbox{subject to}\qquad}\nu_{i} \geq0,\qquad i = 1, \ldots, n.\nonumber
\end{eqnarray}
Let $\bolds\beta_{k+1}$ denote a minimizer of problem~(\ref
{lqs-lp-almost-20}). This leads to an iterative optimization procedure
as described in Algorithm~\ref{algoseq-LO1}.

%alg1
\begin{algorithm}[t]
\caption{Sequential linear optimization algorithm for the LQS problem}\label{algoseq-LO1}
\begin{longlist}[1.]
\item[1.] Initialize with $\bolds\beta_{1}$, and for $k \geq1$
perform the following steps 2--3 for a predefined tolerance parameter ``Tol.''
\item[2.] Solve the linear optimization problem~(\ref{lqs-lp-almost-20})
%with objective function $Q\left( \left(\B\nu, \theta, \B\beta\right)
and let $(\bolds\nu_{k+1}, \theta_{k+1}, \bolds\beta_{k
+1})$ denote a minimizer.
\item[3.] If
$  ( \llvert y_{(q)} - \mathbf{x}'_{(q)}\bolds\beta_{k}\rrvert  -
\llvert y_{(q)} - \mathbf{x}'_{(q)}\bolds\beta_{k+1}\rrvert   ) \leq
\mathrm{Tol}\cdot\llvert y_{(q)} - \mathbf{x}'_{(q)}\bolds\beta
_{k}\rrvert  $
%$F(\B\nu_{k+1}, \theta_{k+1}, \B\beta_{k +1}) - F(\B\nu_{k},
%$f_{q}( \M{r}_{k},\B\beta_{k}) - f_{q}( \M{r}_{k+1},\B\beta_{k+1})
exit; else go to step 2.
\end{longlist}
\end{algorithm}

We next study the convergence properties of Algorithm~\ref{algoseq-LO1}.

\subsubsection*{Convergence analysis of Algorithm~\protect\ref{algoseq-LO1}}\label{secset-up-3}
In representation~(\ref{eqmedian2}), we replace $H_{q}(\bolds
\beta)$ by its dual representation~(\ref{lp-duality-1}) to obtain
%
%e3.7 #&#
\begin{eqnarray}\label{lqs-lp-almost-1}
&& f_{q}(\bolds\beta):= \min_{\bolds{\nu}, \theta}\qquad F(\bolds\nu, \theta, \bolds\beta):= \theta (n - q + 1 ) + \sum
_{i=1}^{n} \nu_{i} - H_{q+1}(\bolds\beta)\nonumber
\\
&&\qquad\mbox{subject to}\qquad \theta+ \nu_{i} \geq\bigl\llvert
y_{i} - \mathbf {x}'_{i}\bolds\beta\bigr\rrvert, \qquad i = 1, \ldots,n,
\\
&&\phantom{\qquad\mbox{subject to}\qquad} \nu_{i} \geq0, \qquad i = 1, \ldots, n.\nonumber
\end{eqnarray}
Note that the minimum of problem~(\ref{lqs-lp-almost-1})
%
%e3.8 #&#
\begin{eqnarray}
\label{lqs-lp-almost-1-equi}
&& \min_{\bolds{\nu}, \theta, \bolds\beta}\qquad F(\bolds \nu, \theta, \bolds\beta)\nonumber
\\
&&\qquad\mbox{subject to}\qquad \theta+ \nu_{i} \geq\bigl\llvert
y_{i} - \mathbf {x}'_{i}\bolds\beta\bigr\rrvert,\qquad i = 1, \ldots,n,
\\
&&\phantom{\qquad\mbox{subject to}\qquad}\nu_{i} \geq0,\qquad i = 1, \ldots, n\nonumber
\end{eqnarray}
equals to $\min_{\bolds\beta} f_{q}(\bolds\beta)$, which
is also the minimum of~(\ref{eqmedian-q}), that is,
$\min_{\bolds\beta} f_{q}(\bolds\beta) = \min_{\bolds\beta} \llvert  r_{(q)}\rrvert  $.
The objective function $F(\bolds\nu, \theta, \bolds\beta
)$ appearing in~(\ref{lqs-lp-almost-1}) is the sum of a linear
function in $(\bolds\nu, \theta)$ and a concave function in
$\bolds\beta$ and the constraints are convex.

%Note that the minimization Problem~\eqref{lqs-lp-almost-1-equi} is
%nonconvex.
%Note that the following function:
% Consider a linearization of the concave function $-H_{q}(\B\beta)
Note that the function
\begin{eqnarray}\label{major-1-Hfn}
&& Q\bigl((\bolds\nu, \theta, \bolds
\beta); \bar{\bolds\beta}\bigr)
\nonumber\\[-8pt]\\[-8pt]\nonumber
&&\qquad = \theta( n-q+1 ) + \sum
_{i=1}^{n} \nu_{i} - \bigl\langle\partial
H_{q+1}(\bar{\bolds\beta}), \bolds \beta- \bar{\bolds\beta}
\bigr\rangle- H_{q+1}(\bar{\bolds\beta}),\nonumber
\end{eqnarray}
which is linear in the variables $(\bolds\nu, \theta,
\bolds\beta)$ is a linearization of $F(\bolds\nu, \theta, \bolds\beta)$ at the
point~$\bar{\bolds\beta}$. Since $H_{q+1}(\bolds
\beta)$ is convex in $\bolds\beta$, the function
$Q((\bolds\nu, \theta, \bolds\beta); \bar{\bolds\beta})$
is a majorizer of $F(\bolds\nu, \theta, \bolds\beta)$
for \emph{any} fixed $\bar{\bolds\beta}$ with equality
holding at $\bar{\bolds\beta}= \bolds\beta$, that is,
\[
Q\bigl((\bolds\nu, \theta, \bolds\beta); \bar{\bolds\beta}\bigr) \geq F(
\bolds\nu, \theta, \bolds \beta) \qquad \forall\bolds\beta\quad\mbox{and}\quad Q\bigl((
\bolds\nu, \theta, \bar{\bolds\beta}); \bar{\bolds\beta}\bigr) = F(
\bolds\nu, \theta, \bar{\bolds\beta}). %
\]
Observe that problem~(\ref{lqs-lp-almost-20}) minimizes the function
$Q((\bolds\nu, \theta, \bolds\beta); \bolds\beta_{k})$.

It follows that for every fixed $\bar{\bolds\beta}$, an
optimal solution of the following linear optimization problem:
%
%e3.9 #&#
\begin{eqnarray}\label{lqs-lp-almost-2}
&& \min_{\bolds{\nu}, \theta, \bolds\beta}\qquad  Q\bigl((\bolds\nu, \theta, \bolds
\beta); \bar{\bolds\beta}\bigr)\nonumber
\\
&&\qquad\mbox{subject to}\qquad \theta+ \nu_{i} \geq\bigl\llvert
y_{i} - \mathbf {x}'_{i}\bolds\beta\bigr
\rrvert, \qquad i = 1, \ldots,n,
\\
&&\phantom{\qquad\mbox{subject to}\qquad} \nu_{i} \geq0, \qquad i = 1, \ldots, n,\nonumber
\end{eqnarray}
provides an upper bound to the minimum of problem~(\ref
{lqs-lp-almost-1-equi}), and hence the global minimum of the LQS
objective function.
We now define the first-order optimality conditions of problem~(\ref
{lqs-lp-almost-1}).

%de1 #&#
\begin{mydef}\label{defloc-min-1}
A point $(\bolds\nu_*, \theta_*, \bolds\beta_*)$
satisfies the first-order optimality conditions for the minimization
problem~(\ref{lqs-lp-almost-1-equi}) if (a) $(\bolds\nu_*,
\theta_*, \bolds\beta_*)$ is feasible for problem~(\ref
{lqs-lp-almost-1}) and (b) $(\bolds\nu_*, \theta_*, \bolds
\beta_*)$ is a minimizer of the following LO problem:
%
%e3.10 #&#
\begin{eqnarray}\label{stat-F-fn-1}
&& \Delta_{*}:=  \min_{\bolds\nu, \theta, \bolds\beta}\qquad \left\langle\nabla F (\bolds\nu_*, \theta_*, \bolds \beta_*), \left(\matrix{ \bolds\nu-
\bolds\nu_{*}
\vspace*{-1.5pt}\cr
\theta- \theta_{*}
\vspace*{-1.5pt}\cr
\bolds\beta- \bolds
\beta_{*}} \right)\right\rangle\nonumber
\\
&&\qquad\mbox{subject to}\qquad \theta+ \nu_{i} \geq\bigl\llvert
y_{i} - \mathbf {x}'_{i}\bolds\beta\bigr
\rrvert, \qquad i = 1, \ldots,n,
\\
&&\phantom{\qquad\mbox{subject to}\qquad} \nu_{i} \geq0, \qquad i = 1, \ldots, n,\nonumber
\end{eqnarray}
where, $ \nabla F (\bolds\nu_*, \theta_*, \bolds\beta
_*)$ is a subgradient of the function $F (\bolds\nu_*, \theta
_*, \bolds\beta_*)$.
\end{mydef}

It is easy to see that, if $(\bolds\nu_*, \theta_*, \bolds
\beta_*)$ satisfies the first-order optimality conditions as in
Definition~\ref{defloc-min-1}, then
$\Delta_{*} = 0 $.

%
%re1 #&#
\begin{rem}\label{remloc-min-1}
Note that if $(\bolds\nu_*, \theta_*, \bolds\beta_*)$
satisfies the first-order optimality conditions for the minimization
problem~(\ref{lqs-lp-almost-1-equi}), then $\bolds\beta_*$
satisfies the first-order stationarity conditions for the LQS
minimization problem~(\ref{eqmedian-q}).
\end{rem}

Let us define $\Delta_{k}$ as a measure of suboptimality of the tuple
$(\bolds\nu_{k}, \theta_{k}, \bolds\beta_{k})$
from first-order stationary conditions, given in Definition~\ref{defloc-min-1}
%
%e3.11 #&#
\begin{equation}
\label{eqdefdelta-k} \Delta_{k}:= \left\langle\nabla F(\bolds
\nu_{k}, \theta_{k}, \bolds\beta_{k }), \pmatrix{
\bolds\nu_{k+1} - \bolds\nu_{k}
\vspace*{-1.5pt}\cr
\theta_{k+1} -
\theta_{k}
\vspace*{-1.5pt}\cr
\bolds\beta_{k +1} - \bolds\beta_{k}}
\right\rangle,
\end{equation}
where $ \{(\bolds\nu_{k}, \theta_{k}, \bolds\beta
_{k }) \}_{k \geq1}$ are as defined in Algorithm~\ref{algoseq-LO1}.

Note that $\Delta_{k} \leq0$. If $\Delta_{k}=0$, then the point
$(\bolds\nu_{k}, \theta_{k}, \bolds\beta_{k})$ satisfies
the first-order stationary conditions.
If $\Delta_{k}<0$, then we can improve the solution further.
The following theorem presents the rate at which $\Delta_{k}
\rightarrow0$.
%

%th3.1 #&#
\begin{teo}\label{labthm1}
\textup{(a)} The sequence $(\bolds\nu_{k}, \theta_{k}, \bolds
\beta_{k })$ generated by Algorithm~\ref{algoseq-LO1} leads~to a
decreasing sequence of objective values $F(\bolds\nu_{k+1}, 
\theta_{k+1}, \bolds\beta_{k+1}) \leq F(\bolds\nu_{k},\break
\theta_{k}, \bolds\beta_{k}), k \geq1$
that converge to a value $F_*$.

\textup{(b)} The measure of suboptimality $\{\Delta_{k}\}_{K \geq k \geq
1}$ admits a $O(1/K)$ convergence rate, that is,
\[
\frac{ F(\bolds\nu_{1}, \theta_{1}, \bolds\beta_{1}) -
F_* }{K} \geq\min_{k = 1,\ldots,K} ( -\Delta_{k} ),
\]
where $F(\bolds\nu_{k}, \theta_{k}, \bolds\beta_{k})
\downarrow F_*$.

\textup{(c)} As $K \rightarrow\infty$ the sequence satisfies the
first-order stationary conditions as in Definition~\ref{defloc-min-1}
for problem~(\ref{lqs-lp-almost-1-equi}).
\end{teo}

\begin{pf}
Part (a). Since the objective function in~(\ref
{lqs-lp-almost-2}) is a linearization of the concave function~(\ref
{lqs-lp-almost-1}), Algorithm~\ref{algoseq-LO1} leads to a decreasing
sequence of objective values:
\[
f_{q}(\bolds\beta_{k+1})= F(\bolds\nu_{k+1}, \theta
_{k+1}, \bolds\beta_{k +1}) \leq F(\bolds\nu_{k},
\theta _{k}, \bolds\beta_{k})= f_{q}(\bolds
\beta_{k}).
\]
Thus, the sequence $F(\bolds\nu_{k}, \theta_{k}, \bolds
\beta_{k})$ is decreasing and bounded below, hence it converges---we
denote the limit as $F_*$.

Part (b).
We make use of the concavity of $ F(\bolds\nu, \theta,
\bolds\beta)$ which follows since it can be written as the sum
of a linear function in $(\bolds\nu, \theta)$ and $-H_{q+1}
(\bolds\beta)$, which is a
concave function in $\bolds\beta$.
This gives rise to the following inequality:
%
%e3.12 #&#
\begin{eqnarray}\label{bound-1}
&& F(\bolds\nu_{k+1}, \theta_{k+1}, \bolds
\beta_{k +1}) - F(\bolds\nu_{k}, \theta_{k}, \bolds
\beta_{k })
\nonumber\\[-8pt]\\[-8pt]\nonumber
&&\qquad \leq \left\langle\nabla F(\bolds\nu_{k},
\theta_{k}, \bolds \beta_{k }), \pmatrix{ \bolds
\nu_{k+1} - \bolds\nu_{k}
\vspace*{-1.5pt}\cr
\theta_{k+1} -
\theta_{k}
\vspace*{-1.5pt}\cr
\bolds\beta_{k +1} - \bolds\beta_{k}}
\right\rangle.
\end{eqnarray}

Considering inequality~(\ref{bound-1}) for $k = 1, \ldots, K$, the
notation~(\ref{eqdefdelta-k}) and adding up the terms we have
%
%e3.13 #&#
\begin{eqnarray}
&& \sum_{k=1}^{K} \bigl( F(\bolds
\nu_{k}, \theta_{k}, \bolds\beta_{k}) - F(\bolds
\nu_{k+1}, \theta_{k+1}, \bolds\beta_{k+1}) \bigr)
\geq \sum_{k = 1}^{K} ( -
\Delta_{k} ),
\end{eqnarray}
that is,
%e3.14 #&#
\begin{equation}
F(\bolds\nu_{1}, \theta_{1}, \bolds\beta
_{1}) - F(\bolds\nu_{K+1}, \theta_{K+1}, \bolds\beta
_{K+1}) \geq K \Bigl( \min_{k = 1,\ldots, K} ( -
\Delta_{k} ) \Bigr),\label{bound-2-1}
\end{equation}
that is,
%e3.15 #&#
\begin{equation}
\frac{ F(\bolds\nu_{1}, \theta_{1},
\bolds\beta_{1}) - F_* }{K}  \geq \Bigl( \min_{k = 1,\ldots, K} (
-\Delta_{k} ) \Bigr).\label{bound-2-2}
\end{equation}

In the above, while moving from line~(\ref{bound-2-1}) to~(\ref
{bound-2-2}) we made use of the fact that
$F(\bolds\nu_{K+1}, \theta_{K+1}, \bolds\beta_{K+1})
\geq F_*$, where the decreasing sequence $F(\bolds\nu_{k},
\theta_{k}, \bolds\beta_{k})$
converges to $F_*$. Equation~(\ref{bound-2-2}) provides a convergence
rate for the algorithm.

Part (c).
As $K \rightarrow\infty$, we see that
$\Delta_{k} \rightarrow0$---corresponding to the first-order
stationarity condition~(\ref{stat-F-fn-1}). This also
corresponds to a local minimum of~(\ref{eqmedian-q}).

This completes the proof of the theorem.
\end{pf}

%s3.2 #&#
\subsection{A first-order subdifferential based algorithm for the LQS problem}\label{secsubgrad1}
Subgradient descent methods have a long history in nonsmooth convex
optimization~[\citet{shor1985minimization,nesterov2004introductory}].
If computation of the subgradients turns out to be inexpensive, then
subgradient based methods are quite effective in
obtaining a moderate accuracy solution with relatively low
computational cost. For nonconvex and nonsmooth functions,
a subgradient need not exist, so the notion of a subgradient needs to
be generalized.
For nonconvex, nonsmooth functions having certain regularity properties
(e.g., Lipschitz functions) subdifferentials exist and form a natural
generalization of
subgradients~[\citet{clarke1990optimization}].
Algorithms based on subdifferential information oracles [see,
e.g.,~\citet{shor1985minimization}]
are thus used as natural generalizations of subgradient methods for nonsmooth,
nonconvex optimization problems.
While general subdifferential-based methods can become quite
complicated based on appropriate choices
of the subdifferential and step-size sequences, we
propose a simple subdifferential based method for approximately
minimizing $f_{q}(\bolds\beta)$ as we describe below.
Recall that $f_{q}(\bolds\beta)$ admits a representation as the
difference of two simple convex functions of the form~(\ref{eqmedian2}).
It follows that $f_{q}(\bolds\beta)$ is Lipschitz~[\citet{rock-conv-96}],
almost everywhere differentiable and any element belonging to the set difference
\[
\partial f_{q}(\bolds\beta) \in{\large{\bolds\partial }}
{H}_{q}(\bolds\beta) - {\large{\bolds\partial}} {H}_{q+1}(
\bolds\beta),
\]
where, ${\large{\bolds\partial}} {H}_{m}(\bolds\beta)$
(for $m = q, q+1$) is the set of subgradients
defined in~(\ref{subdiff-top-r-conv});
is a \emph{subdifferential}~[\citet{shor1985minimization}] of
$f_{q}(\bolds\beta)$.

In particular, the quantity
\[
\partial f_{q}(\bolds\beta) = -\sgn\bigl(y_{(q)} - \mathbf
{x}'_{(q)}\bolds\beta\bigr)\mathbf{x}_{(q)}
\]
is a subdifferential of the function $f_{q}(\bolds\beta)$ at
$\bolds\beta$.

%The general methods can become quite complicated based on the choice
%of the subgradient and step-size sequence.
Using the definitions above, we propose a first-order subdifferential based
%subdifferential based method and not
%a subgradient method. Furthermore, the method does not necessarily
%lead to a deceasing sequence. However, we will adhere to ``subgradient
%descent" for easier terminology.}
method for the LQS problem as described in Algorithm~\ref{algoGD}, below.

%alg2
\begin{algorithm}[t]
\caption{Subdifferential based algorithm for the LQS problem}\label{algoGD}
\begin{longlist}[1.]
\item[1.] Initialize $\bolds\beta_{1}$, for $\operatorname{MaxIter}
\geq k\geq1$ do the following:
\item[2.] $\bolds\beta_{k+1} = \bolds\beta_{k} - \alpha_{k}
\partial f_{q}(\bolds\beta_{k})$
where $\alpha_{k}$ is a step-size.
\item[3.] Return $\min_{1 \leq k \leq\operatorname{MaxIter}}
f_{q}(\bolds\beta_{k})$ and
$\bolds\beta_{k^*}$ at which the minimum is attained, where $k^*
= \argmin_{1 \leq k \leq\operatorname{MaxIter}} f_{q}(\bolds
\beta_{k})$.
\end{longlist}
\end{algorithm}
While various step-size choices are possible, we found the following
simple fixed step-size sequence to be quite useful in our experiments:
\[
\alpha_{k} = \frac{1}{\max_{i = 1, \ldots, n} \llVert   \mathbf{x}_{i}\rrVert  _2},
\]
where, the quantity $\max_{i = 1, \ldots, n} \llVert   \mathbf{x}_{i}\rrVert  _2$
may be interpreted as an upper bound to the subdifferentials
of $f_{q}(\bolds\beta)$. Similar constant step-size based rules
are often used in subgradient descent methods for convex optimization.

% This choice of step-size is often used in subgradient based methods
%(with fixed step-sizes) for nonsmooth convex optimization.
%Consider the problem of minimizing a convex (nonsmooth) function $G(\B
%simple subgradient method applied to this problem is given by:
%where $ \partial G(\B\beta_{k})$ is a subdifferential and $\alpha_{k}$
%is the step-size. Different choices of step-sizes lead to
%different methods. The subdifferential method is not a descent method.
%In practice, the convergence may be slow, but the
%cost per iteration being cheap, it is used as a useful tool for
%large-scale problems. Motivated by the idea of subgradient methods in
%convex optimization we use subgradient methods for the nonconvex
%nonsmooth Problem~\eqref{eqmedian-q-abs}. The use of
%subgradient (or more generally Clarke differential) methods are not
%new in the optimization literature---see for example the
%classical works of~\cite{} and the more recent works of~\cite{}.
%The algorithm outlined above is
%not a descent sequence, so we typically take the $\B\beta_{K}$ for
%which the objective value is the minimum.
%Local search based methods to optimally select the step-size length
%based on one dimensional optimization leads

%Note that this algorithm is very simple, has low cost per iteration
%and with several initializations seems to provide a good local
%solution.

%s3.3 #&#
\subsection{A hybrid algorithm}\label{sechybrid-1}
%The subgradient method described in Section~\ref{secsubgrad1}, with
%different starting points,
%aggressively ``explores'' different regions of the LQS objective
%function containing good local minima.
Let $\hat{\bolds\beta}_{\mathrm{GD}}$ denote the estimate
produced by Algorithm~\ref{algoGD}.
Since Algorithm~\ref{algoGD}
runs with a fixed step-size, the estimate $\hat{\bolds\beta
}_{\mathrm{GD}}$ need not be a local minimum of the LQS problem.
Algorithm~\ref{algoseq-LO1}, on the other hand, delivers an estimate
$\hat{\bolds\beta}_{\mathrm{LO}}$, say,
which\vspace*{2pt} is a \emph{local} minimum of the LQS objective function. We
found that if
$\hat{\bolds\beta}_{\mathrm{GD}}$ obtained from the
subdifferential method is used as a warm-start for the sequential
linear optimization algorithm,
the estimator obtained improves upon $\hat{\bolds\beta
}_{\mathrm{GD}}$
in terms of the LQS objective value. This leads to the proposal of a
hybrid version of Algorithms~\ref{algoseq-LO1}~and~\ref{algoGD}, as
presented in Algorithm~\ref{algohybrid} below.

%alg3
\begin{algorithm}[b]
\caption{A hybrid algorithm for the LQS problem}\label{algohybrid}
\begin{longlist}[1.]
\item[1.] Run Algorithm~\ref{algoGD} initialized with $\bolds\beta
_{1}$ for $\operatorname{MaxIter}$ iterations.
Let $\hat{\bolds\beta}_{\mathrm{GD}}$ be the solution.
\item[2.] Run Algorithm~\ref{algoseq-LO1} with $\hat{\bolds
\beta}_{\mathrm{GD}}$ as the initial solution and Tolerance parameter
``Tol'' to obtain
$\hat{\bolds\beta}_{\mathrm{LO}}$.

\item[3.] Return $\hat{\bolds\beta}_{\mathrm{LO}}$ as the
solution to Algorithm~\ref{algohybrid}.
\end{longlist}
\end{algorithm}

%s3.4 #&#
\subsection{Initialization strategies for the algorithms}\label{secinit-strategies}
Both Algorithms~\ref{algoseq-LO1}~and~\ref{algoGD} are
sensitive to initializations $\bolds\beta_{1}$.
We run each algorithm for a prescribed number of runs ``RUNS'' (say),
and consider the solution that gives the best objective value among them.
For the initializations, we found two strategies to be quite useful.

\subsubsection*{Initialization around LAD solutions}
One\vspace*{1pt} method is based on the LAD solution, that is, $\hat{\bolds\beta}{}^{(\mathrm{LAD})}$ and random initializations around $\hat {\bolds\beta}{}^{(\mathrm{LAD})}$
given by $ [\hat{\beta}^{(\mathrm{LAD})}_{i} - \eta\llvert \hat{\beta}^{(\mathrm{LAD})}_{i}\rrvert, \hat{\beta}^{(\mathrm{LAD})}_{i} +
\eta\llvert \hat{\beta}^{(\mathrm{LAD})}_{i}\rrvert   ]$, for
$i = 1, \ldots, p$, where $\eta$ is a predefined \mbox{number} say $\eta\in
\{ 2, 4\}$. This initialization strategy leads to $\bolds\beta
^{1}$, which we denote by the ``LAD'' initialization.
%The source of randomization in this method lies in the neighborhood of
%the
%LAD solution.
\subsubsection*{Initialization around Chebyshev fits}
% \label{secinit-strategies-cheb}
Another initialization strategy is inspired by a geometric
characterization of the LQS solution [see~\citet{Stromberg1993}
and also Section~\ref{secprops-lqs}].
Consider a subsample $\mathcal J \subset\{1, \ldots, n\}$ of size of
size $(p+1)$ and the associated
$\ell_\infty$ regression fit (also known as the Chebyshev fit) on the
subsample $(y_{i}, \mathbf{x}_{i}), i \in\mathcal J$ given by
\[
\hat{\bolds\beta}_{\mathcal J} \in\argmin_{\bolds\beta
} \Bigl( \max
_{i \in\mathcal J} \bigl\llvert y_{i} - \mathbf
{x}_{i}'\bolds\beta\bigr\rrvert \Bigr).
\]
Consider a number of random subsamples $\mathcal J$ and the associated
coefficient-vector $\hat{\bolds\beta}_{\mathcal J}$ for every
$\mathcal J$.
The estimate $\hat{\bolds\beta}_{\mathcal J^*}$ that produces
the minimum value of the LQS objective function is taken as
$\bolds\beta_{1}$.
We denote $\bolds\beta_{1}$ chosen in this fashion as the best
Chebyshev fit or ``Cheb'' in short.
%The source of randomness in this strategy lies in the
%subsamples drawn to compute the Chebyschev fit.

Algorithm~\ref{algohybrid}, in our experience, was found to be less
sensitive to initializations.
Experiments demonstrating the different strategies described above are
discussed in Section~\ref{seccomps-1}.

%It is worth noting that, $\widehat{\B\beta}_{\mbox{LP}}$ is a fixed

%Though this does not interfere with the definition of the $q$th
%quantile,
%Due to this reason, the MIO formulation takes a long time to obtain
%tight lower bounds.

%s4 #&#
\section{Properties of the LQS solutions for arbitrary datasets}\label{secprops-lqs}
In this section, we prove that key properties
of optimal LQS solutions hold without assuming that the data
$(\mathbf{y}, \mathbf{X})$ are in general position as it is done in
the literature to date~[\citet{rousseeuw1984least,rousseeuw2005robust,Stromberg1993}].
For this purpose, we utilize the MIO characterization of the
LQS problem.
Specifically:
\begin{longlist}[(2)]
\item[(1)]
We show in Theorem~\ref{teoexist} that an optimal solution
to the LQS problem (and in particular the LMS problem)
always exists, for \emph{any} $(\mathbf{y}, \mathbf{X})$ and $q$.
The theorem also shows that an optimal LQS solution is given by
the $\ell_{\infty}$ or Chebyshev regression fit to a subsample of
size $q$ from the sample $(y_{i}, \mathbf{x}_{i})$, $i = 1, \ldots, n$,
thereby generalizing the results of \citet{Stromberg1993}, which
require $(\mathbf{y}, \mathbf{X})$ to be in general position.
\item[(2)] We show in Theorem \ref{propkey2} that the absolute values of
some of the residuals are equal to the optimal solution value of the
LQS problem,
without assuming that the data is in general position.
\item[(3)] We show in Theorem~\ref{teo-break-1} a new result that the
breakdown point of the optimal value of the LQS objective is $(n-q+1)/n$
without assuming that the data is in general position. For the LMS
problem $q=n-\lfloor n/2 \rfloor$, which leads to the sample breakdown
point of\vadjust{\goodbreak}
LQS objective of $(\lfloor n/2 \rfloor+1)/n$, independent of the
number of covariates $p$.
In contrast, it is known that LMS solutions have a sample breakdown
point of $(\lfloor n/2 \rfloor-p+2)/n$ (when the data is in general position).
\end{longlist}

%
%th4.1 #&#
\begin{teo}\label{teoexist}
The LQS problem is equivalent to the following:
%
%e4.1 #&#
\begin{equation}
\label{max-l2-norm1} \min_{\bolds\beta} \llvert r_{(q)}\rrvert =
\min_{ {\mathcal I} \in
\Omega_{q} } \Bigl( \min_{\bolds\beta} \llVert
\mathbf{y}_{I} - \mathbf{X}_{I}\bolds\beta\rrVert
_\infty \Bigr),
\end{equation}
%
%where, for a vector $\M{s}$, the norm $\left\| \M{s}\right\| _\infty$ denotes the
%standard $\ell_{\infty}$-norm,
%of a vector $\M{s}$ denotes the maximum absolute value of its
%coordinates,
where, $ \Omega_{q}:= \{ {\mathcal I} \dvtx  {\mathcal I} \subset \{
1, \ldots, n  \}, \llvert  {\mathcal I} \rrvert  = q \} $
and $(\mathbf{y}_{I}, \mathbf{X}_{I})$ denotes the subsample $(y_{i},
\mathbf{x}_{i}), i \in{\mathcal I}$.
\end{teo}

\begin{pf}
Consider the MIO formulation~(\ref{lqs-reg-form-1}) for the LQS problem.
Let us take a vector of binary variables $\bar{z}_{i}\in\{ 0, 1\}$, $i
= 1, \ldots, n$ with $\sum_{i} \bar{z}_{i} = q$,
feasible for problem~(\ref{lqs-reg-form-1}). This vector $\bar{\mathbf{z}}:= (\bar{z}_{1}, \ldots, \bar{z}_{n})$ gives rise to a
subset $\mathcal I \in\Omega_{q}$ given by
\[
\mathcal I = \bigl\{ i  | \bar{z}_{i} = 1, i
\in\{1, \ldots, n \} \bigr\}.
\]
Corresponding to this subset $\mathcal I$ consider the subsample
$(y_{\mathcal I}, \mathbf{X}_{\mathcal I})$ and the associated
optimization problem
%
%e4.2 #&#
\begin{equation}
\label{max-norm-reg-1} T_{\mathcal I} = \min_{\bolds\beta} \llVert
\mathbf{y}_{I} - \mathbf{X}_{I}\bolds\beta\rrVert
_\infty,
\end{equation}
and let $\bolds\beta_{\mathcal I}$ be a minimizer of~(\ref
{max-norm-reg-1}).
Observe that $\bar{\mathbf{z}}, \bolds\beta_{\mathcal I}$ and
$\bar{r}_{i} = y_{i} - \mathbf{x}_{i}'\bolds\beta_{\mathcal
I}$, $i = 1, \ldots, n$
is feasible for
problem~(\ref{lqs-reg-form-1}). Furthermore, it is easy to see that,
if $\mathbf{z}$ is taken to be equal to $\bar{\mathbf{z}}$,
% i.e.
%$z_{i} = \bar{z}_{i}, i = 1, \ldots, n$
then the minimum value of
problem~(\ref{lqs-reg-form-1}) with respect to the variables
$\bolds\beta$ and $r^+_{i}, r^{-}_{i}, \mu_{i}, \bar{\mu
}_{i}$ for $i = 1, \ldots, n$
is given by $\llvert \bar{r}_{(q)}\rrvert  = T_{\mathcal I}$. Since every choice of
$\mathbf{z} \in\{0,1\}^n$ with $\sum_i z_{i} = q$ corresponds to a subset
$\mathcal I \in\Omega_{q}$, it follows that the minimum value of
problem~(\ref{lqs-reg-form-1}) is given by the minimum value of
$T_{\mathcal I}$ as $\mathcal I$ varies over
$\Omega_{q}$.

Note that the minimum in problem~(\ref{max-l2-norm1}) is
attained since it is a minimum over finitely many subsets $\mathcal I
\in\Omega_{q}$.
This shows that an optimal solution to the LQS problem always exists,
without any assumption
on the geometry or orientation of the sample points $(\mathbf{y},
\mathbf{X})$.
This completes the proof of the equivalence~(\ref{max-l2-norm1}).
\end{pf}

%
%
%co1 #&#
\begin{cor}\label{cor-1}
Theorem~\ref{teoexist} shows that an optimal LQS solution for \emph{any} sample $(\mathbf{y}, \mathbf{X})$
is given by the Chebyshev or $\ell_{\infty}$ regression fit to a
subsample of size
$q$ from the $n$ sample points. In particular, for every optimal LQS solution
there is a
${\mathcal I}_* \in\Omega_{q}$ such that
%
%e4.3 #&#
\begin{equation}
\hat{\bolds\beta}{}^{(\mathrm{LQS})} \in\argmin _{\bolds\beta} \llVert
\mathbf{y}_{{\mathcal I}_*} - \mathbf {X}_{ {\mathcal I}_*}\bolds\beta\rrVert
_{\infty}.
\end{equation}
\end{cor}

We next show that, at an optimal solution
of the LQS problem, some of the absolute values of the residuals are
all equal to the minimum objective value of the LQS problem,
generalizing earlier work by \citet{Stromberg1993}.
%We will use the shorthand notation $\mtilde{y} = \M{y}_{ {\mathcal
%I}_*}$, $\mtilde{X}=\M{X}_{ {\mathcal I}_*}$ for convenience.
Note that problem~(\ref{max-norm-reg-1}) can be written as the
following linear optimization problem:
%
%e4.4 #&#
\begin{eqnarray}\label{max-norm-reg-1-eq}
&& \mini_{t, \bolds\beta}\qquad t,
\nonumber\\[-8pt]\\[-8pt]\nonumber
&&\qquad \mbox{subject to}\qquad - t
\leq{y}_{i} - \mathbf{x}'_{i}\bolds\beta\leq t,\qquad
i \in {\mathcal I}_*.
\end{eqnarray}
The Karush Kuhn Tucker (KKT) [\citet{BV2004}] optimality
conditions of problem~(\ref{max-norm-reg-1-eq}) are given by
%
%e4.5 #&#
\begin{eqnarray} \label{line-kkt-1}
\sum_{i \in{\mathcal I}_* } \bigl( \nu^{-}_{i}
+ \nu^{+}_{i} \bigr) &=& 1,\nonumber
\\
\sum_{i \in{\mathcal I}_*} \bigl( \nu^{-}_{i}
- \nu^{+}_{i} \bigr) \mathbf{x}_{i} &=&
\mathbf{0},\nonumber
\\
\nu^{+}_{i} \bigl( {y}_{i} -
\mathbf{x}'_{i}\hat{\bolds\beta} - t^* \bigr) &=&0\qquad \forall i \in{\mathcal I}_*,
\\
\nu^{-}_{i} \bigl( {y}_{i} -
\mathbf{x}'_{i}\hat{\bolds\beta} + t^* \bigr) &=& 0\qquad \forall i \in{\mathcal I}_*,\nonumber
\\
\nu^{+}_{i}, \nu^{-}_{i} &\geq& 0\qquad\forall i \in{\mathcal I}_*,\nonumber
\end{eqnarray}
where $\hat{\bolds\beta}, t^*$ are optimal
solutions\footnote{We use the shorthand $\hat{\bolds\beta
}$ in place of $\hat{\bolds\beta}{}^{(\mathrm{LQS})}$.}
to~(\ref{max-norm-reg-1-eq}).
%$$ t + \nu^{+}_{i} ( \widetilde{y}_{i} - \M{x}'_{i}\B\beta- t ) -

Let us denote
%
%e4.6 #&#
\begin{equation}
\label{eqistar} %{\mathcal I}^{+}:= \{ i \left| i \in{\mathcal I}_*, \nu^+_{i} >0
% \mbox{or} \nu^{-}_{i} >0 \},
{\mathcal I}^{+}:= \bigl\{ i | i
\in{\mathcal I}_*, \bigl\llvert \nu^+_{i} - \nu^{-}_{i}
\bigr\rrvert >0 \bigr\},
\end{equation}
clearly, on this set of indices at least one of $\nu^+_{i}$ or $\nu
^-_{i}$ is nonzero, which implies
$\llvert  {y}_{i} - \mathbf{x}'_{i}\hat{\bolds\beta}\rrvert  = t^*$.
This gives the following bound:
\[
\bigl\llvert {\mathcal I}^{+} \bigr\rrvert \leq \bigl\llvert \bigl\{
i \in { \mathcal I}_* \dvtx  \bigl\llvert {y}_{i} - \mathbf{x}'_{i}
\hat {\bolds\beta} \bigr\rrvert = t^* \bigr\} \bigr\rrvert. %
\]
It follows from~(\ref{line-kkt-1}) that $\llvert  {\mathcal I}^{+}
\rrvert  > \rnk ([\mathbf{x}_{i}, i \in{\mathcal I}^{+}]
 )$. We thus have
%Using the above along with~\eqref{line-kkt-1} it follows that
%
\[
\bigl\llvert \bigl\{ i \in{ \mathcal I}_* \dvtx  \bigl\llvert {y}_{i} -
\mathbf {x}'_{i}\hat{\bolds\beta}\bigr\rrvert = t^*
\bigr\} \bigr\rrvert \geq \bigl\llvert {\mathcal I}^{+} \bigr\rrvert >
\rnk \bigl(\bigl[\mathbf{x}_{i}, i \in{\mathcal I}^{+}\bigr]
\bigr).
\]
In particular, if the $\mathbf{x}_{i}$'s come from a continuous distribution
%and $\left| {\mathcal I}^{+}\left|\geq p$
then with probability one:
\[
\rnk \bigl(\bigl[\mathbf{x}_{i}, i \in{\mathcal I}^{+}
\bigr] \bigr) = p\quad\mbox{and}\quad \bigl\llvert \bigl\{ i \in{ \mathcal I}_* \dvtx
\bigl\llvert {y}_{i} - \mathbf {x}'_{i}
\hat{\bolds\beta}\bigr\rrvert = t^* \bigr\} \bigr\rrvert \geq(p +1).
\]
This leads to the following theorem.

%th4.2 #&#
\begin{teo} \label{propkey2}
Let ${\mathcal I}_* \in\Omega_{q}$ denote a subset of size $q$ which
corresponds to an optimal LQS solution (see Corollary~\ref{cor-1}).
Consider the KKT optimality conditions of the Chebyshev fit to this
subsample $(\mathbf{y}_{{\mathcal I}_{*}}, \mathbf{X}_{{\mathcal
I}_*})$ as given
by~(\ref{line-kkt-1}). Then
\[
\bigl\llvert \bigl\{ i \in{ \mathcal I}_* \dvtx  \bigl\llvert {y}_{i} -
\mathbf {x}'_{i}\hat{\bolds\beta}\bigr\rrvert = t^*
\bigr\} \bigr\rrvert \geq \bigl\llvert {\mathcal I}^{+} \bigr\rrvert >
\rnk \bigl(\bigl[\mathbf{x}_{i}, i \in{\mathcal I}^{+}\bigr]
\bigr),
\]
where $\hat{\bolds\beta},{\mathcal I}^{+}$ are as defined
in~(\ref{line-kkt-1}) and~(\ref{eqistar}).
\end{teo}

%s4.1 #&#
\subsection{Breakdown point and stability of solutions}
In this section, we revisit the notion of a breakdown point of
estimators and derive sharper results for the problem without the assumption
that the data is in general position.
%We briefly review the definition of the finite sample breakdown point
%of an estimator.
%Donoho and Huber \cite{} introduced the finite sample notion of
%breakdown point of an estimator $\Theta(\M{y}, \M{X})$.
%Suppose the original sample is $(\M{y}, \M{X})$ and $m$ of the sample
%points have been replaced arbitrarily---let
%$\M{y} + \Delta_{\M{y}}, \M{X} + \Delta_{\M{X}})$ denote the perturbed
%sample. Let
%$$\alpha(m; \Theta; (\M{y}, \M{X}))=\sup_{\Delta} \left\|  \Theta((\M{y},
%denote the maximal change in the estimator under this perturbation.
%The finite sample breakdown point of the estimator $\Theta$ is defined
%as follows:
%$$\varepsilon( \Theta; (\M{y}, \M{X}) ):=\min_{m} \{ \frac{m}{n}
% \left| \alpha(m; \Theta; (\M{y}, \M{X})) = \infty\} $$
%Recall the definition of the finite sample breakdown point of an
%estimator~\eqref{eqbreakdown-est1},\eqref{eqbreakdown-est2}.
Let $\Theta(\mathbf{y}, \mathbf{X})$ denote an estimator based on a
sample $(\mathbf{y}, \mathbf{X})$.
Suppose the original sample is $(\mathbf{y}, \mathbf{X})$ and $m$ of
the sample points have been replaced arbitrarily---let
$(\mathbf{y} + \Delta_{\mathbf{y}}, \mathbf{X} + \Delta_{\mathbf
{X}})$ denote the perturbed sample. Let
%
%e4.7 #&#
\begin{equation}
\label{eqbreakdown-est1} \alpha\bigl(m; \Theta; (\mathbf{y}, \mathbf{X})\bigr)=\sup
_{(\Delta
_{\mathbf{y}}, \Delta_{\mathbf{X}} )} \bigl\llVert \Theta(\mathbf{y}, \mathbf{X}) -\Theta(
\mathbf{y} + \Delta_{\mathbf{y}}, \mathbf{X} + \Delta_{\mathbf{X}}) \bigr
\rrVert,
\end{equation}
denote the maximal change in the estimator under this perturbation,
where $\llVert  \cdot\rrVert  $ denotes the standard Euclidean norm.
The finite sample breakdown point of the estimator $\Theta$ is defined
as follows:
%
%e4.8 #&#
\begin{equation}
\label{eqbreakdown-est2} \eta\bigl( \Theta; (\mathbf{y}, \mathbf{X}) \bigr):=\min
_{m} \biggl\{ \frac{m}{n} \bigg| \alpha\bigl(m; \Theta; (
\mathbf{y}, \mathbf {X})\bigr) = \infty \biggr\}.
\end{equation}

%estimator, i.e., $\widehat{\B\beta}^{(\mbox{LMS})}$
%is given by $(\lfloor n/2 \rfloor- p + 2)/n$ under the assumption
%that $p>1$ and the sample points are in general position.
%We show that this is a conservative estimate of the breakdown point,
%it is possible to obtain a better estimate by resorting to the MIO
%formulation of the LQS problem.

%Suppose we consider the following modified breakdown point of an
%estimator $\Theta$ as follows. Instead of considering the
%change in the value of the estimator, we consider the change in the
%value of the objective function, i.e.we define
%$$\widetilde{\alpha}(m; f; (\M{y}, \M{X}))=\sup_{\Delta} \left\|  f(
%where $\Theta((\M{y}, \M{X})) \in\argmin_{\Theta} f( \Theta; (\M{y}
%,\M{X}) ).$

We will derive the breakdown point of the minimum value of the LQS
objective function, that is,
$\llvert r_{(q)}\rrvert  = \llvert y_{(q)} - \mathbf{x}'_{(q)} \hat{\bolds\beta}{}^{(\mathrm{LQS})}\rrvert $, as defined in~(\ref{eqmedian2}).
%%where $\B\beta= (\B\beta^{(\mathrm{LQS})})$ i.e. the minimum value
%of the LQS objective function~\eqref{eqmedian-q-abs}.

%th4.3 #&#
\begin{teo}\label{teo-break-1}
Let $ \hat{\bolds\beta}{}^{(\mathrm{LQS})}$ denote an
optimal solution and
$\Theta:=\Theta(\mathbf{y}, \mathbf{X})$ denote the optimum
objective value to the LQS problem
% i.e., $\left|r_{(q)}\left| $ where, $r_{i} = y _{i} - \M{x}_{i}' \widehat{\B
for a given dataset $(\mathbf{y}, \mathbf{X})$,
where the $(y_{i}, \mathbf{x}_{i})$'s are not necessarily in general position.
Then the finite sample breakdown point of $\Theta$ is $(n - q + 1) / n$.
\end{teo}

\begin{pf}
We will first show that the breakdown point of $\Theta$ is strictly
greater than $(n - q) /n$.
Suppose we have a corrupted sample $(\mathbf{y} + \Delta_{\mathbf
{y}}, \mathbf{X} + \Delta_{\mathbf{X}})$, with $m = n - q $ replacements
in the original sample. Consider the equivalent LQS formulation~(\ref{max-l2-norm1}) and let ${\mathcal I}_{0}$ denote the unchanged
sample indices.
Consider the inner convex optimization
problem appearing in~(\ref{max-l2-norm1}), corresponding to the index
set ${\mathcal I}_{0}$:
%
%e4.9 #&#
\begin{equation}
\label{eqn-maxmin-1} T_{{\mathcal I}_{0} } (\mathbf{y} + \Delta_{\mathbf{y}}, \mathbf{X}
+ \Delta_{\mathbf{X}}) = \min_{\bolds\beta} \llVert \mathbf
{y}_{{\mathcal I}_{0}} - \mathbf{X}_{{\mathcal I}_{0}} \bolds \beta\rrVert
_\infty,
\end{equation}
with\vspace*{1pt} $\bolds\beta_{ {\mathcal I}_{0} } (\mathbf{y} + \Delta
_{\mathbf{y}}, \mathbf{X} + \Delta_{\mathbf{X}})$ denoting a minimizer
%and $T_{{\mathcal I}_{0} } (\M{y} + \Delta_{\M{y}}, \M{X} + \Delta_{
of the convex optimization problem~(\ref{eqn-maxmin-1}).
Clearly, both a minimizer and the minimum objective value are finite
and neither depends upon $(\Delta_{\mathbf{y}}, \Delta_{\mathbf{X}})$.
Suppose
\[
T_{{\mathcal I}_*}(\mathbf{y} + \Delta_{\mathbf{y}}, \mathbf{X} +
\Delta_{\mathbf{X}}) = \min_{{\mathcal I} \in\Omega_{q}} T_{\mathcal I}(\mathbf{y}
+ \Delta_{\mathbf{y}}, \mathbf{X} + \Delta_{\mathbf{X}})
\]
denotes the minimum value of the LQS objective function corresponding
to the perturbed sample, for some ${\mathcal I}_* \in\Omega_{q}$,
then it follows that: $T_{{\mathcal I}_*}(\mathbf{y} + \Delta
_{\mathbf{y}}, \mathbf{X} + \Delta_{\mathbf{X}}) \leq T_{{\mathcal
I}_{0} } (\mathbf{y} + \Delta_{\mathbf{y}}, \mathbf{X} + \Delta
_{\mathbf{X}})$---which clearly implies that
% bounded-ness of the optimal value of the LQS objective function in
%presence of perturbation.
%Hence, if $\B\beta_{{\mathcal I}^*}(\M{y}, \M{X})$ is a solution to
%the LQS problem then
the quantity $\llVert  T_{{\mathcal I}_{0} } (\mathbf{y} + \Delta_{\mathbf
{y}}, \mathbf{X} + \Delta_{\mathbf{X}}) - \Theta\rrVert  $ is bounded above
and the bound does not depend upon $ (\Delta_{\mathbf{y}}, \Delta
_{\mathbf{X}} )$. This shows that the breakdown point of $\Theta$ is
strictly larger than $\frac{(n - q)}{n}$.\vadjust{\goodbreak}

% $\left\|  \B\beta_{{\mathcal I}^*} - \B\beta_{ {\mathcal I}_{0} } \right\| $ is
%bounded and does not depend upon $(\Delta_{\M{y}}, \Delta_{\M{X}})$.
%This implies that $\varepsilon( \Theta; (\M{y}, \M{X}) ) > \frac{(n -
%q)}{n}$.
We will now show that the breakdown point of the estimator is less than
or equal to $(n - q+1)/n$.
If the number of replacements is given by $m = n - q + 1$, then
it is easy to see that every ${\mathcal I} \in\Omega_{q}$ includes a
sample $i_{0}$ (say) from the replaced sample units.
If $ (\delta_{y_{i_0}}, \delta'_{\mathbf{x}_{i_{0}}})$ denotes\vspace*{1pt} the
perturbation corresponding to the $i_0$th sample, then it is easy to
see that
\[
T_{\mathcal I}(\mathbf{y} + \Delta_{\mathbf{y}}, \mathbf{X} +
\Delta_{\mathbf{X}}) \geq\bigl\llvert (y_{i} - \mathbf{x}_{i_{0}}
\bolds \beta_{\mathcal I}) + \bigl(\delta_{y_{i_0}} -
\delta'_{\mathbf
{x}_{i_{0}}}\bolds\beta_{\mathcal I}\bigr) \bigr\rrvert,
\]
where $ \bolds\beta_{\mathcal I}$ is a minimizer for the
corresponding problem~(\ref{eqn-maxmin-1}) (with ${\mathcal I}_{0} =
{\mathcal I}$).
It is possible to choose $\delta_{y_{i_0}}$
%in the direction of
%$\sgn(y_{i} - \M{x}_{i_{0}}\B\beta_{\mathcal I})$ with $ \delta'_{
such that the r.h.s. of the above inequality becomes arbitrarily
large. Thus, the finite-sample breakdown point of the estimator $\Theta
$ is $\frac{(n - q)+1}{n}$.
\end{pf}

For the LMS problem $q=n-\lfloor n/2 \rfloor$, which leads to the
sample breakdown point of $\Theta$ of $(\lfloor n/2 \rfloor+1)/n$,
independent of the number of covariates $p$.
In contrast, LMS solutions have a sample breakdown point of $(\lfloor
n/2 \rfloor-p+2)/n$. In other words, the optimal solution value
is more robust than optimal solutions to the LMS problem.

\section{Computational experiments}\label{seccomps-1}
In this section, we perform computational experiments demonstrating the
effectiveness of our algorithms in terms of quality of solutions
obtained, scalability and
speed.

All computations were done in MATLAB version {\texttt R2011a} on a
64-bit linux machine, with 8 cores and 32 GB RAM.
For the MIO formulations we used {\textsc{Gurobi}} [\citet{gurobi}] via its MATLAB interface.

We consider a series of examples including synthetic and real-world
datasets showing that our proposed methodology
consistently finds high quality solutions of problems of sizes up to $n
={}$10,000 and $p=20$.
For moderate-large sized examples, we observed that
global optimum solutions are obtained usually within a few minutes (or
even faster), but it takes longer to deliver a certificate of global optimality.
%The MIO formulation~\eqref{lqs-reg-form-1} delivers solutions with a
%certificate of global optimality for problems of
%sizes $n = 501$ in a very reasonable amount of time, the times taken
%to obtain tight global lower bounds
%when $n$ is of the order of thousands with $p \geq10$ can be longer.
Our continuous optimization based methods
enhance the performance of the MIO formulation, the margin of
improvement becomes more significant with increasing problem sizes.
In all the examples, there is an appealing common theme---if the MIO
algorithm is terminated early, the procedure provides a bound on its
suboptimality.

In Section~\ref{synthetic-egs}, we describe the synthetic datasets
used in our experiments.
%Section~\ref{secdeeper-1} studies the performances of Algorithms~
%datasets.
Section~\ref{secdeeper-1} presents a deeper understanding of
Algorithms~\ref{algoseq-LO1},~\ref{algoGD} and~\ref{algohybrid}.
Section~\ref{seccompare-methods} presents comparisons of
Algorithms~\ref{algoseq-LO1},~\ref{algoGD} and~\ref{algohybrid} as
well as the MIO algorithm
with state of the art algorithms for the LQS. In Section~\ref{secreal-data}, we illustrate the performance of our algorithms on
real-world data sets.
Section~\ref{secglobal-opt} discusses the evolution of lower bounds
and global convergence certificates for the problem.
Section~\ref{seclarge-scale-method1} describes scalability
considerations for larger problems.

%In all the cases, however, the MIO formulation~\eqref{lqs-reg-form-1}
%along
%with our continuous optimization based methods obtains the global
%minimum
%consistently delivers global optimal solutions for problems-sizes up
%to $n = 75, p =3$ and $n \approx50, p = 7$---if the algorithms are
%terminated before
%global convergence, the MIO formulation provides lower bounds
%certifying its distance from global optimality. For smaller problems,
%%trim option's parameter order\dvtx  left bottom right top
%%\includegraphics[trim = 10mm 80mm 20mm 5mm, clip, width=3cm]{chick}
%%%%%%
%% about to comment
%%trim option's parameter order\dvtx  left bottom right top
%%%\begin{comment}
%%%\begin{figure}%
%%%\caption{Figure showing the typical evolution of the MIO formulation~
%%%[Top-row] Alcohol dataset with $n = 44, q = 31$ with $p=5$ (left
%panel) and
%%%$p = 7$ (right panel).
%%%[Middle row] HBK dataset warm-started with the least squares
%solution, for $q=60$ (left panel) and $q=45$ (right panel).
%%%[bottom row] HBK dataset warm-started with Algorithm~\protect
%%%\end{figure}
%%%\end{comment}

%s5.1 #&#
\subsection{Synthetic examples}\label{synthetic-egs}
We considered a set of synthetic examples,
following~\citet{Rousseeuw2006CLR11170811117088}.
We generated the model matrix $\mathbf{X}_{n \times p}$ with i.i.d.
Gaussian entries $N(0,100)$ and took $\bolds\beta\in\Re^{p}$
to be a vector of all ones.
Subsequently, the response is generated as $\mathbf{y} = \mathbf
{X}\bolds{\beta} + \bolds\varepsilon$, where $\varepsilon
_{i}\sim\mathrm{N}(0,10)$, $i = 1, \ldots, n$.
Once $(\mathbf{y}, \mathbf{X})$ have been generated, we corrupt a
certain proportion $\pi$ of the sample in two different ways:
\begin{longlist}[(A)]
\item[(A)] $\lfloor\pi n \rfloor$ of the samples are chosen at
random and the first coordinate of the data matrix $\mathbf{X}$, that is,
$x_{1j}$'s are replaced by $x_{1j}\leftarrow x_{1j} + 1000$.

\item[(B)] $\lfloor\pi n \rfloor$ of the samples are chosen at
random out of which the covariates of half of the points are changed as
in item~(A);
for the remaining half of the points the responses are corrupted as
$y_{j} \leftarrow y_{j} + 1000$. In this set-up, outliers are added in
\emph{both}
the covariate and response spaces.
\end{longlist}

We considered seven different examples for different values of $(n, p,
\pi)$:
\begin{description}
\item[Moderate-scale:] We consider four moderate-scale examples Ex-1--Ex-4:
\begin{enumerate}
\item[Ex-1:] Data is generated as per (B) with $(n, p, \pi)= (201, 5, 0.4)$.
\item[Ex-2:] Data is generated as per (B) with $(n, p, \pi)= (201,
10, 0.5)$.
\item[Ex-3:] Data is generated as per (A) with $(n, p, \pi)= (501, 5, 0.4)$.
\item[Ex-4:] Data is generated as per (A) with $(n, p, \pi)= (501,
10, 0.4)$.
\end{enumerate}
\item[Large-scale:] We consider three large-scale examples, Ex-5--Ex-7:
\begin{enumerate}
\item[Ex-5:] Data is generated as per (B) with $(n, p, \pi)= (2001,
10, 0.4)$.
\item[Ex-6:] Data is generated as per (B) with $(n, p, \pi)= (5001,
10, 0.4)$.
\item[Ex-7:] Data is generated as per (B) with $(n, p, \pi)= (10,001, 20, 0.4)$.
\end{enumerate}
\end{description}
%
%Among the above, we will consider the four examples Ex-1--Ex-4 to be
%moderate scale problems and Ex--5--Ex--7 to be large problems.

%s5.2 #&#
\subsection{A deeper understanding of 
Algorithms~\texorpdfstring{\protect\ref{algoseq-LO1}}{1}, 
\texorpdfstring{\protect\ref{algoGD}{2}}
and~\texorpdfstring{\protect\ref{algohybrid}}{3}}\label{secdeeper-1}
For each of the synthetic examples Ex-1--Ex-4, we compared the
performances of the different continuous optimization based algorithms
proposed in this paper---Algorithms~\ref{algoseq-LO1}, \ref{algoGD} and~\ref{algohybrid}. For each of the
Algorithms~\ref{algoseq-LO1},~\ref{algoGD}, we considered two
different initializations, following the strategy described in
Section~\ref{secinit-strategies}:
\begin{longlist}[(Cheb)]
\item[(LAD)] This is the initialization from the LAD solution, with
$\eta= 2$ and number of random initializations taken to be 100. This is
denoted in Table~\ref{table-gran-1-small} by the moniker ``LAD.''

%
%t1 #&#
\begin{sidewaystable}%[!h]
\tablewidth=\textwidth
\tabcolsep=0pt
\caption{Table showing performances of different continuous
optimization based methods proposed in this paper, for examples, Ex-1--Ex-4.
For every example, the top row ``Accuracy'' is Relative Accuracy
[see~(\protect\ref{rel-accu})] and
the numbers inside parenthesis denotes standard errors (across the
random runs); the lower row denotes the time taken (in cpu seconds).
Results are averaged over 20 different random instances of the problem.
Algorithm~\protect\ref{algohybrid} seems to be the clear winner among
the different examples, in terms of the quality
of solutions obtained.
Among all the algorithms considered, Algorithm~\protect\ref{algohybrid} seems to be least sensitive to initializations}\label{table-gran-1-small}
\begin{tabular*}{\tablewidth}{@{\extracolsep{\fill}}@{}lccccccc@{}}
\hline
\multirow{2}{71pt}{\break \break \textbf{Example} $\bolds{(n,p,\pi)}$\break \textbf{q}} & & \multicolumn{6}{c@{}}{\textbf{Algorithm used}}\\[-6pt]
& & \multicolumn{6}{c@{}}{\hrulefill}\\
& & \multicolumn{2}{c}{\textbf{Algorithm~\ref{algoseq-LO1}}} & \multicolumn{2}{c}{\textbf{Algorithm~\ref{algoGD}}} & \multicolumn{2}{c@{}}{\textbf{Algorithm~\ref{algohybrid}}}\\[-6pt]
& & \multicolumn{2}{c}{\hrulefill} & \multicolumn{2}{c}{\hrulefill} & \multicolumn{2}{c@{}}{\hrulefill}\\
& & \textbf{(LAD)} & \textbf{(Cheb)} & \textbf{(LAD)} & \textbf{(Cheb)} & \textbf{(LAD)} & \textbf{(Cheb)}\\
\hline
Ex-1 (201, 5, 0.4) & Accuracy & 49.399 (2.43) & 0.0 (0.0) & 0.233 (0.03)& 0.240 (0.02) & 0.0 (0.0) & 0.0 (0.0)\\
$q=121$ & Time (s) & \phantom{0}24.05 & \phantom{0}83.44 & 3.29 & 83.06 & \phantom{0}36.13 & 118.43
\\[3pt]
Ex-2 (201, 10, 0.5)& Accuracy &43.705 (2.39) & 5.236 (1.73) & 1.438 (0.07) & 1.481 (0.10) & 0.0 (0.0) & 0.0 (0.0) \\
$q=101$ & Time (s) & \phantom{0}54.39 & 133.79 & 3.22 & 73.14 & \phantom{0}51.89 & 125.55
\\[3pt]
Ex-3 (501, 5, 0.4) & Accuracy & 2.897 (0.77) & 0.0 (0.0) & 0.249 (0.05) & 0.274 (0.06) & 0.0 (0.0) & 0.0 (0.0)\\
$q=301$ & Time (s) & \phantom{0}83.01 & 158.41 & 3.75 & 62.36 & 120.90 & 179.34
\\[3pt]
Ex-4 (501, 10, 0.4) & Accuracy & 8.353 (2.22) & 11.926 (2.31) & 1.158 (0.06) & 1.083 (0.06) & 0.0 & 0.0 \\
$q=301$ & Time (s) &192.02 & 240.99 & 3.76 & 71.45 & 155.36 & 225.09 \\
\hline
\end{tabular*}
\end{sidewaystable}

\item[(Cheb)] This is the initialization from the Chebyshev fit. For
every initialization, forty different subsamples were taken to estimate
$\bolds\beta_{1}$, 100 different initializations were
considered. This method is denoted by the moniker ``Cheb'' in
Table~\ref{table-gran-1-small}.
\end{longlist}
Algorithm~\ref{algoseq-LO1}, initialized at the ``LAD'' method
(described above) is denoted by Algorithm~\ref{algoseq-LO1} (LAD), the
same notation
carries over to the other remaining combinations of Algorithms~\ref
{algoseq-LO1} and \ref{algoGD} with initializations ``LAD'' and ``Cheb.''
Each of the methods
Algorithms~\ref{algoGD} (LAD) and~\ref{algoGD} (Cheb), lead
to an initialization for Algorithm~\ref{algohybrid}---denoted by
Algorithms~\ref{algohybrid} (LAD) and~\ref{algohybrid}
(Cheb), respectively.

In all the examples, we set the MaxIter counter for Algorithm~\ref
{algoGD} at 500 and took the step-size sequence as described in
Section~\ref{secsubgrad1}.
The tolerance criterion ``Tol'' used in Algorithm~\ref{algoseq-LO1}
(and consequently Algorithm~\ref{algohybrid}), was set to $10^{-4}$.

%For each of the above two cases we have a corresponding case for
%Algorithm~\ref{algohybrid}---denoted by
%Algorithm~\ref{algohybrid}(LAD) and Algorithm~\ref{algohybrid}(Cheb)
%respectively.
Results comparing these methods are
summarized in Table~\ref{table-gran-1-small}. To compare the different
algorithms in terms of the quality of solutions obtained, we do the following.
For every instance, we run all the algorithms and obtain the best
solution among them, say, $f_*$. If
$f_{\mathrm{alg}}$ denotes the value of the LQS objective function for
algorithm ``alg,'' then we define the relative accuracy of the solution
obtained by ``alg'' as
%
%e5.1 #&#
\begin{equation}
\label{rel-accu} \mbox{Relative Accuracy} = (f_{\mathrm{alg}} -
f_{*})/f_{*} \times 100.
\end{equation}

To obtain the entries in Table~\ref{table-gran-1-small}, the relative accuracy
is computed for every algorithm (six in all: Algorithms~\ref{algoseq-LO1}---\ref{algohybrid}, two types for each ``LAD'' and
``Cheb'') for every random problem instance corresponding to a
particular example type; and the results are averaged (over 20 runs).
The times reported for Algorithms~\ref{algoseq-LO1} (LAD) and~\ref{algoseq-LO1} (Cheb) includes the times taken to perform
the LAD and Chebyshev fits, respectively.
The same thing applies to Algorithms~\ref{algoGD} (LAD) and~\ref{algoGD} (Cheb).
For Algorithm~\ref{algohybrid} (Cheb) [resp., Algorithm~\ref{algohybrid} (LAD)],
the time taken equals the time taken by Algorithm~\ref{algoGD} (Cheb)
[resp., Algorithm~\ref{algoGD} (LAD)]
and the time taken to perform the Chebyshev (resp., LAD) fits.

In Table~\ref{table-gran-1-small}, we see that Algorithm~\ref{algoGD}
(LAD) converges quite quickly in all the examples. The quality of the
solution, however, depends upon
the choice of \mbox{$p$---}for $p=10$ the algorithm converges to a lower
quality solution when compared to $p=5$. The time till convergence for
Algorithm~\ref{algoGD} is less sensitive to the problem
dimensions---this is in contrast to the other algorithms, where
computation times show a monotone trend depending upon the sizes of $(n,p)$.
Algorithm~\ref{algoGD} (Cheb) takes more time than Algorithm~\ref{algoGD} (LAD), since it spends a
considerable amount of time in performing multiple Chebyshev fits (to
obtain a good initialization). Algorithm~\ref{algoseq-LO1} (LAD) seems
to be sensitive to the
type of initialization used; Algorithm~\ref{algoseq-LO1} (Cheb) is
more stable and it appears that the multiple Chebyshev initialization
guides Algorithm~\ref{algoseq-LO1} (Cheb) to higher quality solutions.
Algorithm~\ref{algohybrid} (both variants) seem to be the clear winner
among the various algorithms---this does not come as a surprise since,
intuitively it aims
at combining the \emph{best features} of its constituent algorithms.
Based on computation times,
Algorithm~\ref{algohybrid} (LAD) outperforms Algorithm~\ref
{algohybrid} (Cheb),\vadjust{\goodbreak} since it avoids the computational overhead of
computing several Chebyshev fits.

%
%t2 #&#
\begin{table}%[ht]
\tabcolsep=0pt
\caption{Table showing performances of various algorithms for the LQS
problem for different moderate-scale examples as described in the text.
For each example, ``Accuracy'' is Relative Accuracy [see~(\protect\ref{rel-accu})],
the numbers within brackets denote the standard errors;
the lower row denotes the averaged cpu time (in secs) taken by the
algorithm. All results are averaged over 20 random examples.
The {MIO formulation}~(\protect\ref{lqs-reg-form-1}) warm-started
with {Algorithm}~\protect\ref{algohybrid} seems to be the best
performer in terms of obtaining the best solution.
The combined time taken by {MIO formulation}~(\protect\ref{lqs-reg-form-1}) (warm-start) and {Algorithm}~\protect\ref{algohybrid}
(which is used as a warm-start) equals the run-time
of {MIO formulation}~(\protect\ref{lqs-reg-form-1}) (cold-start)}\label{table-notgran-1-small}
\begin{tabular*}{\tablewidth}{@{\extracolsep{\fill}}@{}lc cccc@{}}
\hline
\multirow{2}{71pt}{\break \break \textbf{Example} $\bolds{(n,p,\pi)}$\break \textbf{q}} & & \multicolumn{4}{c@{}}{\textbf{Algorithm used}}\\[-6pt]
& & \multicolumn{4}{c@{}}{\hrulefill}\\
& & \multirow{3}{32pt}{\centering{\textbf{LQS}\break \textbf{(MASS)}}} &  & \multicolumn{2}{c@{}}{\textbf{MIO formulation~(\ref{lqs-reg-form-1})}}\\[-6pt]
& & & & \multicolumn{2}{c@{}}{\hrulefill}\\
& &  &  \textbf{Algorithm \ref{algohybrid}} & \textbf{(Cold-start)} & \textbf{(Warm-start)}\\
\hline
Ex-1 (201, 5, 0.4) &Accuracy &24.163 (1.31) & 0.0 (0.0) & 60.880 (5.60) & 0.0 (0.0) \\
$q=121$ & Time (s) & 0.02 & \phantom{0}36.13 & \phantom{0}71.46 & \phantom{0}35.32 \\[3pt]
Ex-2 (201, 10, 0.5) &Accuracy & 105.387 (5.26) & 0.263 (0.26) & 56.0141 (3.99) & 0.0 (0.0) \\
$q=101$ & Time (s) & 0.05 & \phantom{0}51.89 & 193.00 & 141.10\\[3pt]
Ex-3 (501, 5, 0.4) & Accuracy & 9.677 (0.99) & 0.618 (0.27) & 11.325 (1.97) & 0.127 (0.11) \\
$q=301$ & Time (s) & 0.05 &120.90 & 280.66 & 159.76\\[3pt]
Ex-4 (501, 5, 0.4) &Accuracy &29.756 (1.99) & 0.341 (0.33) & 27.239 (2.66) & 0.0 (0.0) \\
$q=301$ & Time (s) & 0.08 & 155.36 & 330.88 & 175.52 \\
\hline
\end{tabular*}
\end{table}

%
%s5.3 #&#
\subsection{Comparisons: Quality of the solutions obtained}\label{seccompare-methods}
In this section, we shift our focus from studying the detailed dynamics
of Algorithms~\ref{algoseq-LO1}---\ref{algohybrid}; and compare
the performances of Algorithm~\ref{algohybrid} (which seems to be the
best among the algorithms we propose in the paper),
the MIO formulation~(\ref{lqs-reg-form-1}) and state-of-the art
implementations of the LQS problem as
implemented in the popular \texttt{R}-package \texttt{MASS} (available from
\texttt{CRAN}).
For the MIO formulation~(\ref{lqs-reg-form-1}), we considered two
variations: MIO formulation~(\ref{lqs-reg-form-1}) (cold-start), where
the MIO algorithm is not provided
with any advanced warm-start and MIO formulation~(\ref
{lqs-reg-form-1}) (warm-start), where the MIO algorithm is provided with
an advanced warm-start obtained by Algorithm~\ref{algohybrid}.
The times taken by MIO formulation~(\ref{lqs-reg-form-1}) (warm-start)
do not include the times taken by Algorithm~\ref{algohybrid}, the
combined times are similar to the times
taken by MIO formulation~(\ref{lqs-reg-form-1}) (cold-start).

The focus here is on comparing the quality of upper bounds to the LQS problem.
We consider the same datasets used in Section~\ref{secdeeper-1} for
our experiments.
The results are shown in Table~\ref{table-notgran-1-small}. We see
that MIO formulation~(\ref{lqs-reg-form-1}) (warm-start) is the clear
winner among all the examples,
Algorithm~\ref{algohybrid} comes a close second. MIO formulation~(\ref
{lqs-reg-form-1}) (cold-start) in the absence of advanced warm-starts
as provided by Algorithm~\ref{algohybrid} requires more time to obtain
high quality upper bounds. The state-of-the art algorithm for LQS
(obtained from the \texttt{R}-package \texttt{MASS}) delivers a solution very quickly,
but the solutions obtained are quite far from the global minimum.

%s5.4 #&#
\subsection{Performance on some real-world datasets}\label{secreal-data}
We considered a few real-world datasets popularly used in the context
of robust statistical estimation, as available from the {\texttt R}-package \texttt{robustbase} [\citet
{robust-base-manual,robust-base-paper}].
We used the ``Alcohol'' dataset~(available from the same package),
which is aimed at studying
the solubility of alcohols in water to understand alcohol transport in
living organisms. This dataset
contains physicochemical characteristics of $n = 44$
aliphatic alcohols and measurements on seven numeric variables:
SAG solvent accessible surface-bounded molecular volume ($x_{1}$),
logarithm of the octanol-water partitions coefficient ($x_{2}$), polarizability
($x_{3}$), molar refractivity ($x_{4}$), mass ($x_{5}$), volume ($x_{6}$)
and the response ($y$) is taken to be the logarithm of the solubility.
%%V, logpc, mass, rm, sag
We consider two cases from the Alcohol dataset---the first one has $n =
44$, $p = 5$ where the five covariates were $x_{1},x_{2}, x_{4}, x_{5}, x_{6}$;
the~second example has all the six covariates and an intercept term,
which leads to $p = 7$. We used the MIO formulation~(\ref
{lqs-reg-form-1}) (cold-start) for both the cases.
%found the algorithm to reach a global solution in a very reasonable
%amount of time.
The evolution of the MIO (with upper and lower bounds) for the two
cases are shown in
Figure~\ref{fig-alco-data1}. As expected, the time taken for the
algorithm to converge is larger for $p=7$ than for $p=5$.\vadjust{\goodbreak}

%was used
%without any specialized warm-start.
%seemed to be quite efficient

\begin{figure}%[!h]

\includegraphics{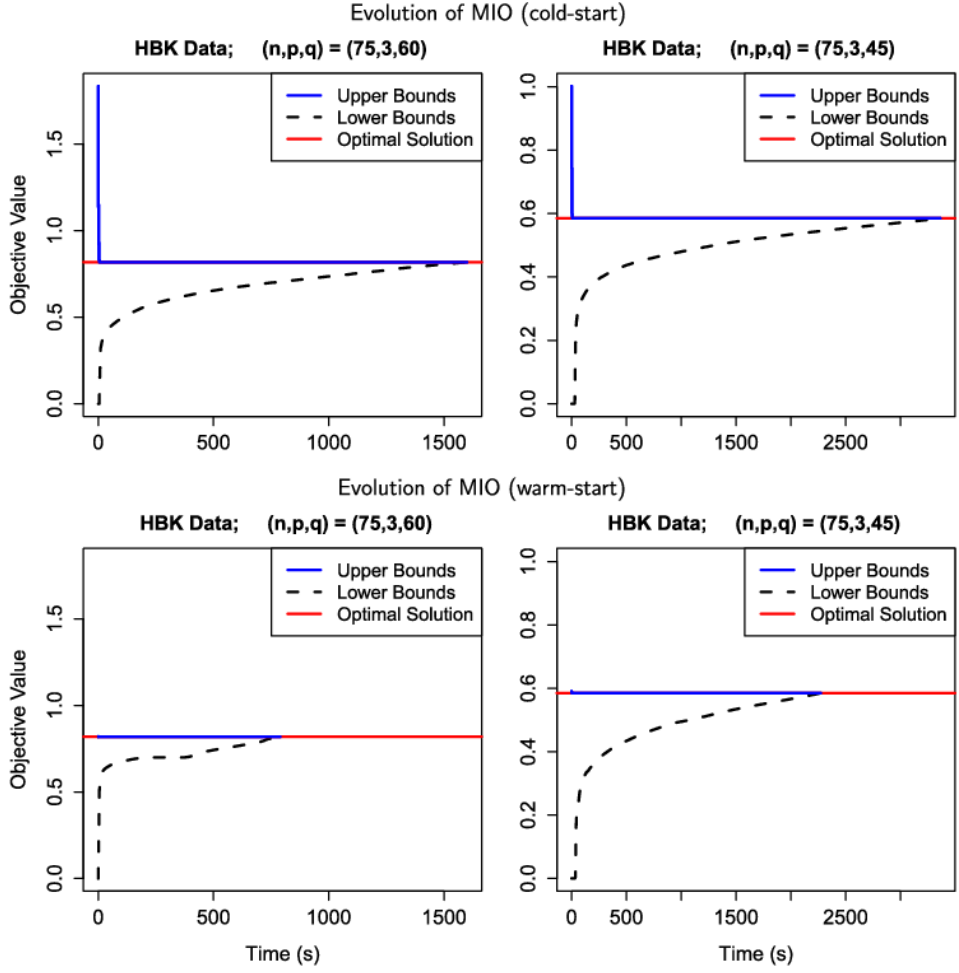}

%{ { \large{ \sf Evolution of MIO (cold-start) } } }\\
%{ { \large\sf Evolution of MIO (warm-start) } } \\
\caption{Figure showing the evolution of the MIO formulation~(\protect\ref{lqs-reg-form-1}) for the HBK dataset with different values of
$q$ with and without warm-starts.
(Top row) MIO formulation warm-started with the least squares solution
(which we denote by ``cold-start''), for $q=60$ (left panel) and $q=45$
(right panel).
(Bottom row) MIO formulation warm-started with Algorithm~\protect\ref{algohybrid} for $q=60$ (left panel) and $q=45$ (right panel).}\label{fig-hbk-data1}
\end{figure}

We considered a second dataset created by~\citet
{hawkins1984location} and available from the \texttt{R}-package
{\texttt{robustbase}}. The dataset consists of 75
observations in four dimensions (one response and three explanatory
variables), that is, $n = 75, p = 3$.
%and provides a good
%example of the masking effect.
We computed the LQS estimate for this example for $q \in\{60, 45\}$.
We used both
the MIO formulation~(\ref{lqs-reg-form-1}) (cold-start) and MIO
formulation~(\ref{lqs-reg-form-1}) (warm-start) and observed that the
latter showed superior convergence speed to global optimality (see
Figure~\ref{fig-hbk-data1}). As expected, the time taken for
convergence was found to increase with decreasing
$q$-values. The results are shown in Figure~\ref{fig-hbk-data1}.

%
%f2 #&#

%s5.5 #&#
\subsection{Certificate of lower bounds and global optimality}\label{secglobal-opt}

\begin{figure}%[!h]

\includegraphics{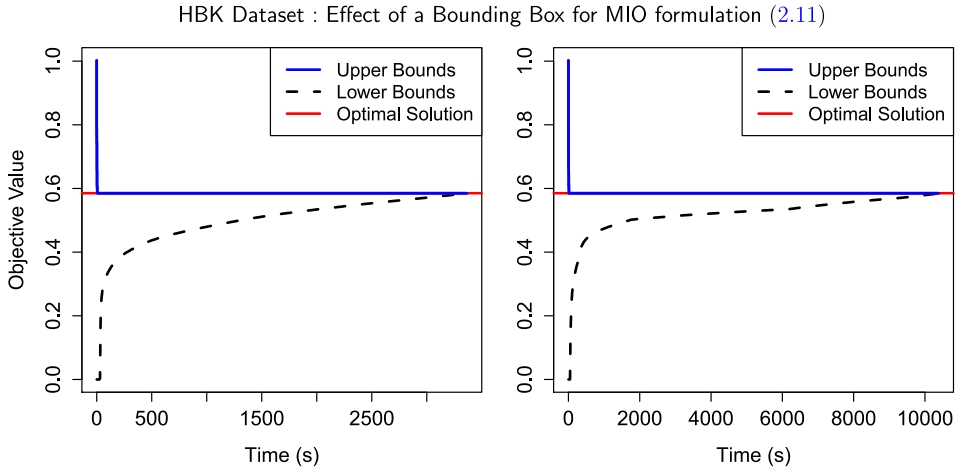}

%{ \large\sf HBK Dataset: Effect of a Bounding Box for MIO
%formulation~\eqref{lqs-reg-form-1}}
\caption{Figure showing the effect of the bounding box for the
evolution of the MIO formulation~(\protect\ref{lqs-reg-form-1}) for
the HBK dataset, with
$(n,p,q) = (75, 3, 45)$. The left panel considers a bounding box of
diameter 6 and the right panel considers a bounding box of diameter 80
centered around the least squares solution.}\label{fig-hbk-data-fraction}
\end{figure}

The MIO formulation~(\ref{lqs-reg-form-1}) for the LQS problem
converges to the global solution. With the aid of advanced MIO
warm-starts as
provided by Algorithm~\ref{algohybrid} the MIO obtains a very high
quality solution very quickly---in most of the examples the solution
thus obtained, indeed turns out to be the global minimum. However, the
certificate of global optimality comes later as the lower bounds of the
problem ``evolve'' slowly; see, for example, Figures~\ref{fig-alco-data1} and~\ref{fig-hbk-data1}.
We will now describe a regularized version of the MIO formulation,
which we found to be quite useful in speeding up the convergence of the
MIO algorithm without any loss in the
accuracy of the solution.
The LQS problem formulation does not contain any explicit
regularization on $\bolds\beta$, it is rather implicit (since
$\hat{\bolds\beta}{}^{(\mathrm{LQS})}$ will be generally bounded).
We thus consider the following modified version of the LQS
problem~(\ref{eqmedian-q}):
%
%e5.2 #&#
\begin{eqnarray}\label{eqmedian-q-bound}
&& \mini_{\bolds\beta}\qquad \llvert r_{(q)}\rrvert,
\nonumber\\[-8pt]\\[-8pt]\nonumber
&&\qquad\mbox{subject to}\qquad \llVert \bolds\beta- \bolds\beta_{0}\rrVert _\infty\leq M
\end{eqnarray}
for some predefined $\bolds\beta_{0}$ and $M\geq0$.
If $\hat{\bolds\beta}_{M}$ solves problem~(\ref
{eqmedian-q-bound}), then it is the global minimum
of the LQS problem in the $\ell_{\infty}$-ball $\{ \bolds\beta
\dvtx  -M\mathbf{1} \leq\bolds\beta- \bolds\beta_{0} \leq
M\mathbf{1} \}$.
In particular, if $\hat{\bolds\beta}{}^{\mathrm{LQS}}$ is
the solution to problem~(\ref{eqmedian-q}), then by
choosing $\bolds\beta_{0}= \mathbf{0}$ and $M \geq\llVert   \hat {\bolds\beta}{}^{\mathrm{LQS}} \rrVert  _\infty$ in~(\ref
{eqmedian-q-bound}); both problems~(\ref{eqmedian-q}) and~(\ref
{eqmedian-q-bound})
will have the same solution. The MIO formulation of problem~(\ref
{eqmedian-q-bound}) is a very simple modification of~(\ref
{lqs-reg-form-1}) with additional box-constraints on
$\bolds\beta$ of the form $ \{ \bolds\beta\dvtx  -M\mathbf{1}
\leq\bolds\beta- \bolds\beta_{0} \leq M\mathbf{1} \}$.
Our empirical investigation suggests that the MIO formulation~(\ref{lqs-reg-form-1}) in presence of box-constraints\footnote{Of course, a
very large value of $M$ will render the box-constraints to be ineffective.}
produces tighter lower bounds than the unconstrained MIO
formulation~(\ref{lqs-reg-form-1}), for a given time limit.
As an illustration of formulation~(\ref{eqmedian-q-bound}), see
Figure~\ref{fig-hbk-data-fraction}, where we use the MIO
formulation~(\ref{lqs-reg-form-1}) with box constraints.
We consider two cases corresponding to $M \in\{ 3, 40\}$; in both the
cases we took $\bolds\beta_{0} = \hat{\bolds\beta}{}^{(\mathrm{LS})}=(0.08,-0.36,0.43)$.
Both these boxes (which are in fact, quite large, given that $\llVert
\hat{\bolds\beta}{}^{\mathrm{(LS)}}\rrVert  _\infty= 0.43$)
contains the (unconstrained)
global solution for the problem. As the figure shows, the evolution of
the lower bounds of the
MIO algorithm toward the global optimum depends upon the radius of the box.

%%% begin figure
%
%f3 #&#
%
%%% end figure
%

%

\begin{figure}%[!h]

\includegraphics{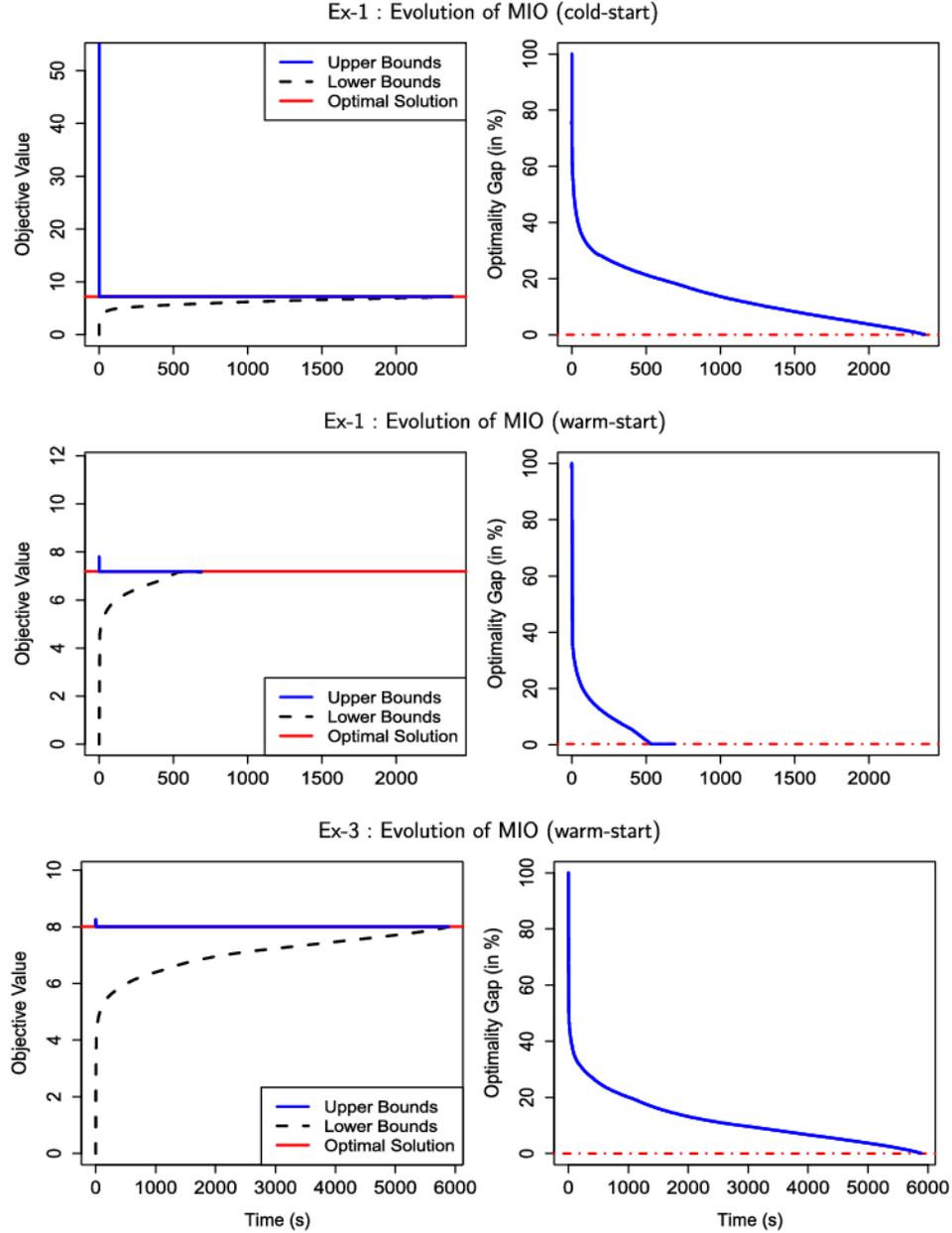}

%{ \large{ \sf Ex-1: Evolution of MIO (cold-start) } }\\
%{\large{ \sf Ex-1: Evolution of MIO (warm-start) } }\\
%{ \large{ \sf Ex-3: Evolution of MIO (warm-start) } }\\
\caption{Figure showing evolution of MIO in terms of upper/lower
bounds (left panel) and Optimality gaps (in \%) (right panel).
Top and middle rows display an instance of Ex-1 with $(n, p, q) = (201,
5, 121)$ with different initializations, that is, MIO~(\protect\ref{lqs-reg-form-1}) (cold-start) and
MIO~(\protect\ref{lqs-reg-form-1}) (warm-start), respectively.
Bottom row considers an instance of Ex-3 with $(n, p, q) = (501, 5,
301)$.}\label{figmio-moderate-p}
\end{figure}

%Though, problems~\eqref{eqmedian-q-bound} are not, in general
%equivalent (unless $M$ is sufficiently large),
We argue that formulation~(\ref{eqmedian-q-bound}) is a more desirable
formulation---the constraint may behave as a regularizer to shrink
coefficients or
if one seeks an unconstrained LQS solution, there are effective ways to
choose to $\bolds\beta_{0}$ and $M$. For example,
if $\bolds\beta_{0}$ denotes the solution obtained by
Algorithm~\ref{algohybrid}, then for $M=\eta\llVert   \bolds\beta_{0}
\rrVert  _{\infty}$, for $\eta\in[1, 2]$ (say),
the solution to~(\ref{eqmedian-q-bound}) corresponds to a global
optimum of the LQS problem inside a box of diameter $2M$ centered at
$\bolds\beta_{0}$.
For moderate sized problems with $n \in\{ 201, 501\}$, we found this
strategy to be useful in certifying global optimality within a
reasonable amount of time.
Figure~\ref{figmio-moderate-p} shows some examples.

%f4 #&#

%s5.6 #&#
\subsection{Scalability to large problems}\label{seclarge-scale-method1}
We present our findings for large-scale experiments performed on
synthetic and real data-set, below.

\subsubsection*{Synthetic large scale examples}
For large-scale problems
with $n \geq5000$ with $p \geq10$, we found that Algorithm~\ref
{algoseq-LO1} becomes computationally demanding due to the
associated LO problems~(\ref{lqs-lp-almost-20}) appearing in step~2 of
Algorithm~\ref{algoseq-LO1}.
On the other hand,
Algorithm~\ref{algoGD} remains computationally inexpensive. So for
larger problems, we propose using a modification of Algorithm~\ref
{algoseq-LO1}---we\vspace*{1pt} run
Algorithm~\ref{algoGD} for several random initializations around the
$\hat{\bolds\beta}{}^{(\mathrm{LAD})}$ solution and find the
best solution among them.
The regression coefficient thus obtained is used
as an initialization for Algorithm~\ref{algoseq-LO1}---we call this
Algorithm~\ref{algohybrid} (large-scale).
Note that this procedure deviates from the vanilla Algorithm~\ref
{algohybrid} (described in Section~\ref{sechybrid-1}), where, we do
\emph{both} steps~1~and~2 for every initialization $\bolds
\beta_{1}$.
For each of the examples Ex-5--Ex-7, Algorithm~\ref{algoGD}
was run for $\operatorname{MaxIter} = 500$, for 100 different
initializations around the LAD solution, the best solution was used as
an initialization for
Algorithm~\ref{algoseq-LO1}. Table~\ref{table-notgran-1-large}
presents the results obtained with
Algorithm~\ref{algohybrid} (large-scale). In addition, the
aforementioned algorithms, Table~\ref{table-notgran-1-large} also presents
MIO (warm-start), that is, MIO formulation~(\ref{lqs-reg-form-1})
warm-started with
Algorithm~\ref{algohybrid} (large-scale) and the LQS algorithm from
the \texttt{R}-package \texttt{MASS}.

\subsubsection*{Large scale examples with real datasets}
In addition to the
above, we considered a large environmental dataset from the \texttt
{R}-package {\texttt{robustbase}}
with hourly measurements of NOx pollution content in the ambient air.
The dataset has $n = 8088$ samples with $p = 4$ covariates (including
the intercept).
The covariates are
square-root of the windspeed $(x_{1})$,
day number ($x_{2}$),
log of hourly sum of NOx emission of cars ($x_{3}$) and intercept,
with response being log of hourly mean of NOx concentration in ambient
air $(y)$.
We considered three different values of $q \in\{ 7279, 6470, 4852 \}$
corresponding to the $90$th, $80$th and $60$th quantile, respectively.
We added a small amount of contamination by changing $\lfloor0.01n
\rfloor$ sample points according to item~(B) in Section~\ref{synthetic-egs}. On the modified dataset,
we ran three different algorithms: Algorithm~\ref{algohybrid}
(large-scale),\footnote{In this example, we initialized Algorithm~\ref
{algoGD} with the best Chebyshev fit from forty different subsamples.
Algorithm~\ref{algoGD} was run for $\operatorname{MaxIter}=500$, with five hundred
random initializations. The best solution was taken as the starting
point of Algorithm~\ref{algohybrid}.} MIO (warm-start), that is, MIO
formulation~(\ref{lqs-reg-form-1}) warm-started with
Algorithm~\ref{algohybrid} (large-scale) and
the LQS algorithm from the \texttt{R}-package \texttt{MASS}. In all the
following cases, the MIO algorithm was run for a maximum of two hours.
We summarize our key findings below:
%t3 #&#
\begin{table}%[!h]
\tabcolsep=0pt
\caption{Table showing performances of various Algorithms for the LQS
problem for different moderate/large-scale examples as described in the text.
For each example, ``Accuracy'' is Relative Accuracy [see~(\protect\ref{rel-accu})]
the numbers within brackets denote the standard errors;
the lower row denotes the averaged cpu time (in secs) taken for the
algorithm. All results are averaged over 20 random examples}\label{table-notgran-1-large}
\begin{tabular*}{\tablewidth}{@{\extracolsep{\fill}}@{}lc cccc@{}}
\hline
\multirow{2}{71pt}{\break \break \textbf{Example} $\bolds{(n,p,\pi)}$\break \textbf{q}} & & \multicolumn{4}{c@{}}{\textbf{Algorithm used}}\\[-6pt]
& & \multicolumn{4}{c@{}}{\hrulefill}\\
& & \multirow{3}{32pt}{\centering{\textbf{LQS}\break \textbf{(MASS)}}} & \multirow{3}{47pt}{\centering{\textbf{Algorithm~\ref{algohybrid} (large-scale)}}}  & \multicolumn{2}{c@{}}{\textbf{MIO formulation~(\ref{lqs-reg-form-1})}}\\[-6pt]
& & & & \multicolumn{2}{c@{}}{\hrulefill}\\
& &  &   & \textbf{(Cold-start)} & \textbf{(Warm-start)}\\
\hline
Ex-5 (2001, 10, 0.4) & Accuracy & 65.125 (2.77) &0.0 (0.0) & 273.543 (16.16) & 0.0 (0.0) \\
$q=1201$ & Time (s) & 0.30 &13.75 & 200& 100\\[3pt]
Ex-6 (5001, 10, 0.4)& Accuracy & 52.092 (1.33) & 0.0 & 232.531 (17.62) & 0.0 (0.0) \\
$q=3001$ & Time (s) & 0.69 & 205.76 & 902 & 450.35 \\[3pt]
Ex-7 (10,001, 20, 0.4)& Accuracy &146.581 (3.77) & 0.0 (0.0) & 417.591 (4.18) & 0.0 (0.0) \\
$q=6001$ &Time (s) & 1.80& 545.88 & 1100 & 550 \\
\hline
\end{tabular*}
\end{table}

\begin{longlist}[(3)]
\item[(1)] For $q = 7279$, the best solution was obtained by MIO
(warm-start) in about 1.6 hours.
Algorithm~\ref{algohybrid} (large-scale) delivered a solution with
relative accuracy~[see~(\ref{rel-accu})] $0.39$\%
in approximately six minutes.
The LQS algorithm from \texttt{R}-package \texttt{MASS}, delivered a solution
with relative accuracy $2.8$\%.
\item[(2)] For $q = 6470$, the best solution was found by MIO (warm-start)
in 1.8 hours.
Algorithm~\ref{algohybrid} (large-scale) delivered a solution with
relative accuracy [see~(\ref{rel-accu})] $0.19$\%
in approximately six minutes.
The LQS algorithm from \texttt{R}-package \texttt{MASS}, delivered a solution
with relative accuracy $2.5$\%.
\item[(3)] For $ q = 4852$, the best solution was found by MIO (warm-start)
in about 1.5 hours.
Algorithm~\ref{algohybrid} (large-scale) delivered a solution with
relative accuracy~[see~(\ref{rel-accu})] $0.14$\%
in approximately seven minutes.
The LQS algorithm from \texttt{R}-package \texttt{MASS}, delivered a solution
with relative accuracy $1.8$\%.
\end{longlist}
Thus, in all the examples above, MIO warm-started with Algorithm~\ref
{algohybrid} (large-scale) obtained the best upper bounds.
Algorithm~\ref{algohybrid} (large-scale) obtained very high quality
solutions, too, but the solutions were all improved by MIO.

%The constraint behaves as a regularizer, and provided shrinkage to the
%$\B\beta$ coefficients.

%%trim option's parameter order: left bottom right top

%s6 #&#
\section{Conclusions}
In this paper, we proposed algorithms for LQS problems based on a
combination of first-order methods from continuous optimization and
mixed integer optimization. Our key conclusions are:
\begin{longlist}[(3)]
\item[(1)] The MIO algorithm with warm start from the continuous
optimization algorithms solves to provable optimality problems of small
($n=100$)
and medium ($n=500$) size problems in under two hours.\vadjust{\goodbreak}
\item[(2)] The MIO algorithm with warm starts finds high quality solutions
for large ($n={}$10,000) scale problems in under two hours outperforming
all state of the art algorithms
that are publicly available for the LQS problem. For problems of this
size, the MIO algorithm does not provide a certificate of optimality in
a reasonable amount of time.
\item[(3)] Our framework enables us to show the existence of an optimal
solution for the LQS problem for \emph{any}
dataset, where the data-points $(y_{i}, \mathbf{x}_{i})$'s are not
necessarily in general position.
Our MIO formulation leads to a simple proof of the breakdown point of
the LQS optimum objective value that holds for general datasets
and our framework can easily incorporate extensions of the LQS
formulation with
polyhedral constraints on the regression coefficient vector.
\end{longlist}


%suskaldyti doi

% imsref loaded by linak, 2014-07-25 10:41:53
%

\printaddresses
\end{document}